\documentclass[acmsmall]{acmart}
\renewcommand\footnotetextcopyrightpermission[1]{}
\fancypagestyle{plain}{
  \fancyfoot{}
  \fancyhead{}
}
\usepackage{enumerate}
\usepackage{enumitem} 
\usepackage{caption}
\usepackage{adjustbox}
\usepackage{graphicx}
\usepackage{subfigure}
\usepackage{subcaption}
\usepackage{wasysym}
\usepackage{booktabs}
\usepackage{longtable}
\usepackage{colortbl}
\usepackage{multirow}
\usepackage[compact]{titlesec}

\usepackage{rotating}
\usepackage{lscape}
\usepackage{color}
\usepackage{ulem}

\AtBeginDocument{%
  }

\begin{document}


\title{Analysing Attacks on Blockchain Systems in a Layer-based Approach}

\author{Joydip Das}
\authornote{Both authors contributed equally to this research.}
\email{joydip007x@gmail.com}
\orcid{0009-0009-6920-5796}
\author{Syed Ashraf Al Tasin}
\authornotemark[1]
\email{ashraf.tasin.000@gmail.com}
\orcid{0009-0004-5573-3452}
\affiliation{%
  \institution{Shahjalal University of Science \& Technology}
  \city{Sylhet 3114}
  \country{Bangladesh}
}

\author{Md. Forhad Rabbi}
\orcid{0000-0002-2468-7973}
\affiliation{%
  \institution{Shahjalal University of Science \& Technology}
  \city{Sylhet 3114}
  \country{Bangladesh}
}
\email{frabbi-cse@sust.edu}

\author{Md Sadek Ferdous}
\orcid{0000-0002-8361-4870}
\affiliation{%
  \institution{BRAC University}
  \city{Dhaka}
  \country{Bangladesh}
  }
\email{sadek.ferdous@bracu.ac.bd}

\renewcommand{\shortauthors}{Das et al.}





\begin{abstract}
Blockchain is a growing decentralized system built for transparency and immutability. There have been several major attacks on blockchain-based systems, leaving a gap in the trustability of this system. This article presents a comprehensive study of 23 attacks on blockchain systems and categorizes them using a layer-based approach. This approach provides an in-depth analysis of the feasibility and motivation of these attacks. In addition, a framework  is proposed that enables a systematic analysis of the impact and interconnection of these attacks, thereby providing a means of identifying potential attack vectors and designing appropriate countermeasures to strengthen any blockchain system.
\end{abstract}
\maketitle

\section{Introduction}
From the very first implementation of blockchain through Bitcoin by Satoshi Nakamoto \citep{originalBitcoin}, this technology has offered improved security and played a vital role in the development and enhancement of many different application domains such as banking sectors, streaming \& copyright services, wallet services, healthcare services, electronic voting, and IoT (Internet of Things). Blockchain's rise can be credited to its idea of efficient ledger management with decentralization and immutability. Unfortunately, even with increased security, blockchain has been shown to be vulnerable in different aspects. Records of security breaches in blockchain-based systems are noticeable with reports claiming a total of \$14.6B worth of funds been stolen, accumulated in 45 countries with 364 incidents, in between 2011 to 2022
\cite{CBA_report_}. In addition, the largest De-Fi (Decentralized Finance) hack to date happened in early 2022, involving more than \$650 million \cite{CBA_report_}. BonqDAO and AllianceBlock experienced a security breach on February 2, 2023, as a result of a flaw in BonqDAO's smart contract. This led to a financial loss of approximately \$120 million \cite{alliance2023}. The biggest financial loss in a 2023 attack occurred in September 23, when an attack on the database of Mixin Network's cloud service provider resulted in a substantial loss of assets on its mainnet, amounting to approximately \$200 million \cite{mixin2023}.

Another evidence of vulnerability can be seen in this publicly revealed data confirming that money lost to blockchain hackers is around \$273 hundreds million with more than 800 such events \cite{SlowMist_report_}. According to data gathered by Comparitech, 6 out of the top 10 expensive crypto attacks occurred in 2021 alone \cite{Comparitech_report_}. Also in the early years of Bitcoin, the famous crypto-exchange network Mt. Gox lost \$474 Million due to security flaws such as transaction mutability \cite{Wired_MtGox_report_}.
Nomad, a crypto start-up, was hacked in early 2022, which was the eighth largest cryptocurrency hack with the damage of \$190 million token \cite{cnbc_nomad_report_}. In 2016, ether worth \$9 billion was taken from the DAO (Decentralized Autonomous Organization) due to some flaws in code \cite{bloomberg_dao_report_}.

These attacks have key impacts on the global economy, with the prediction of the risk of losing \$30 billion per year by 2025 \cite{thePrint_cryptobuzz_report_}. In order to ensure a significant adoption of blockchain, these attacks need to be studied and analyzed for vulnerabilities. In addition, these attacks can be calculated for their impacts, feasibility, and mitigation techniques. Our study aims to tackle the challenge of securing blockchain systems by conducting an in-depth analysis, breaking down the barriers to enhanced protection. Although previous studies have shed light on blockchain-based systems and attack vectors, they often lacked a thorough and systematic analysis of these attacks. Our study seeks to close these gaps by offering an in-depth attack analysis using a layer-based approach, attacker perspective and outcome analysis, and a detailed examination of the connections between attacks across several layers through a custom framework discussed in Section \ref{taxonomy}.

\subsection{Contribution}
This article provides a comprehensive analysis and systematic study of attacks related to different blockchain systems in the context of a four-layer blockchain structure as previously introduced by Ferdous \textit{et al.} in \cite{ferdous2020blockchain}. The primary contributions of this paper are as follows.
\begin{itemize}
    \item We conduct a thorough examination of several security attacks commonly targeted towards different blockchain systems.
    \item For every instance of an attack, we try to find the answers to the following questions.
    \begin{itemize}
        \item Is there a motivation behind these attacks and which specific vulnerabilities are being exploited?
        \item What are the steps in each of these attacks?
        \item What are the potential challenges for attackers in initiating these attacks and what are the potential outcomes?
        \item Can these attacks be considered realistic, and are there any known mitigating strategies?
    \end{itemize}
    In order to answer these questions, we have formulated several criteria, which have been used to analyze each attack.
    \item We investigate the connections between scenarios in which one attack can provide a strategic advantage in executing additional attacks.
    \item Finally, we present a visual summary, in tabular format, of our analysis of each attack. 
\end{itemize}

\subsection{Structure}
The remainder of the article is structured as follows. In Section \ref{chapter2}, we provide the necessary background knowledge of blockchain systems. Next, in Section \ref{RelatedWorks}, we discuss related work relevant to this article. Section \ref{layerBasedAttackAnalysis} gives a comprehensive overview of our framework by discussing the blockchain layers, the taxonomy of properties, and a list of analyzed attacks. Then, in Section \ref{sec:netLayerAttack}, Section \ref{sec:conLayerAttack}, Section \ref{sec:appLayerAttack} and Section \ref{sec:metLayerAttack}, different attacks relevant to each layer are analyzed using the formulated properties. Section \ref{discussion} presents a summary of the findings. Finally, we draw our conclusions in Section \ref{Conclusion}.

\section{Background}
\label{chapter2}
In this section, we provide a brief overview of different aspects of blockchain technology to understand the attacks analyzed in this article.

A peer-to-peer computer network with a distributed ledger is the fundamental idea of blockchain. Cryptographer David Chaum's article `Computer Systems Established, Maintained, and Trusted by Mutually Suspicious Groups' contains the first concept for a blockchain-like technology \cite{8674176}. Later Haber et al. introduced the concept of a collection of blocks linked together using a cryptographic mechanism \cite{haber1991time}. However, the pioneer in making blockchain viable was Satoshi Nakamoto who introduced Bitcoin using a mix of well-established technologies such as a peer-to-peer (P2P) network, digital signature, secure timestamping, and cryptographic hash algorithms \cite{originalBitcoin}. A blockchain is an example of distributed ledger that consists of blocks that are connected to each other with cryptographic mechanisms, thus forming the notion of chain of blocks or blockchain \cite{chowdhury2019comparative}. Each block contains a number of transactions, each transaction representing a transfer of data or value between any two entities. The distributed nature of a blockchain requires a consensus algorithm to ensure that all data stored in a blockchain are synchronized with all P2P nodes. The key innovation of Bitcoin by Satoshi Nakamoto was to achieve a network-wide consensus among the P2P nodes regarding the state of the blockchain in a decentralized way without using any trusted party. All these features enable blockchain to maintain a decentralized, immutable, transparent, efficient, and consensus-based ledger. 

In addition to Bitcoin, Ethereum is another popular blockchain system that offers additional capabilities. Ethereum is a decentralized application platform developed on top of a blockchain \cite{buterin2013ethereum}. Bitcoin is a digital currency, however, Ethereum is more concerned with application development. It accomplishes this by using the notion of smart contracts, which are self-executing code that can automate agreements and transactions. These smart contracts are implemented on a virtual machine, named Ethereum Virtual Machine (EVM), and stored on the blockchain. These contracts can be executed using a transaction that changes the state of the virtual machine. This change of states is also recorded on the Blockchain. Like bitcoin, a distributed consensus algorithm ensures a network-wide agreement over the EVM state and the blockchain data, thus facilitating the notion of immutable code and data. This opens up a wide range of possibilities, including safe marketplaces, complicated financial instruments, and even new types of digital assets. Ethereum features its own coin, Ether (ETH), which is used to pay for different transactions, smart contract execution and data stored on the blockchain.

There are mainly two different types of blockchain as discussed in the following.

\begin{itemize}
    \item Public Blockchain: Public blockchains are decentralized networks that are accessible to anyone who wishes to take part in validating and recording transactions. They offer transparency and accountability as anyone can access the blockchain and verify each of its components. This transparency and decentralization make them ideal for cryptocurrencies and decentralized applications. Bitcoin \cite{BitcoinOriginal} and Ethereum \cite{ethereumHomeEthereumorg} are prime examples of such public blockchain systems.
    
    \item Private Blockchain: Private blockchains are restricted networks where only authorized participants can validate and record transactions. They are suitable for enterprise use cases like supply chain management and financial transactions, where privacy, control and efficient operations are essential. Examples include Hyperledger Fabric \cite{hyperledger} and R3 Corda \cite{r3digitalmarketsCorda}.

\end{itemize}

\subsection{Consensus Mechanism}
Any blockchain system's core component is the underlying consensus algorithm. A consensus algorithm is a fault-tolerant mechanism that is employed to reach agreement on specific decisions or states within a blockchain network. In a blockchain system, consensus is critical since it guarantees that each new block added to the ledger represents the single version of the truth agreed upon by all nodes. Any consensus mechanism consists of these three properties - consistency, availability, and fault tolerance. A consensus algorithm mechanism must guarantee the characteristics of the atomic broadcast (i.e., validity, agreement, integrity, and total order) \cite{ferdous2020blockchain}. Numerous criteria are used to establish the acceptable network condition with consensus. The following section discusses some common consensus algorithms that are employed in various blockchain systems \cite{ferdous2020blockchain}.

\begin{itemize}
  \item \textbf{Proof of Work (PoW):} PoW is based on a simple principle - `A solution that is difficult to find but is easy to verify' \cite{nakamoto2008peer}. PoW involves solving a resource-intensive computational cryptographic puzzle to add new blocks to the blockchain. The PoW mechanism has a difficulty parameter and a node repeatedly solves the cryptographic puzzle to reach that parameter value. If successful, broadcast it to other nodes. The widely implemented version of PoW is based on SHA-256 \cite{concept_pow_basics}. The computer nodes which engage in solving such puzzles are known as miner nodes and the process is known as mining. 
  \item \textbf{Proof of Stake (PoS):} In PoS, miners are known as `Validators'. The network selects a validator through a bidding process. Each validator deposits a portion of their cryptocurrency associated with that network known as stake. The selection of a validator is typically based on this stake. The likelihood of a participant being selected as a validator increases with the amount of cryptocurrency he stakes. However, some PoS systems may also consider additional factors like how long the stakes have been held \cite{king2012peercoin} and randomization \cite{nxt2013whitepaper} to prevent any particular entity from being continuously chosen as a validator. If a validator tries to cheat or use unfair ways and gets detected, he will lose all the stakes deposited previously. PoS is more energy efficient than PoW as it does not consume electricity.  
  \item \textbf{Practical Byzantine Fault Tolerance ( PBFT):} This algorithm was proposed by Castro and Liskov in 1999 \cite{castro1999practical}. In this system, a cluster of replicas process transactions and ultimately creates a new block. The primary replica orders the transaction and gathers approvals from other replicas. Upon receiving enough approvals, the primary replica creates a block and broadcasts it. The system functions properly as long as the proportion of malicious nodes is less than one-third of the total nodes and the primary replica is not compromised. This consensus mechanism is mainly used in private blockchains. 
\end{itemize}

\subsection{Mining Pools and Reward System}
\label{subsec:miningPool}
When miners collaborate to form a sizable collective network for effective mining, it is referred to as a mining pool. Each member of a mining pool contributes computing power to solve a block. If any member finds a block, the entire mining pool is rewarded with the related cryptocurrency. Usually, a pool operator maintains the pool. The pool operator is responsible for the reward distribution and other operational activities \cite{eyal2015miner}. The mining pool reward system is based on `Shares'. A share is a partial block solution. For example, let us assume that a block solution is a number that contains 32 trailing zeros. If a solution with 28 trailing zeros is found, it may be considered as a share or partial proof of work. The share is the main indicator of an individual miner's contribution to the mining pool to find a valid solution. When any participant finds a full proof of work (FPoW), that is, a number with 32 trailing zeros, it is submitted to the pool manager. The pool manager then publishes this FPoW to the blockchain network and the block generation reward is distributed among the participants in one of the many different methods, such as proportional, pay-per-share (PPS) and pay-per-last-N-shares (PPLNS) \cite{rosenfeld2011analysis}. We briefly describe these methods next.

\begin{itemize}
  \item \textbf{Proportional:} This reward scheme is based on rounds. A round is the time interval of finding 2 blocks. In each round, miners keep submitting shares to the pool. If the pool succeeds in finding a block, it gets rewarded and the reward is distributed to the miners by the number of shares they submitted during that round. 
  \item \textbf{Pay-per-share (PPS):} Every miner is immediately rewarded with the expected value of the share’s contribution upon submitting a valid share. The pool operator receives all of the rewards for discovered blocks and pays out miners using the pool's current balance. 
  \item \textbf{Pay-per-last-N-shares (PPLNS):} This scheme is somewhat similar to the `Proportional' scheme. In contrast to Proportional, the miner's payment in this technique is determined based on the last N shares rather than all shares from the previous round. As a result, all miners profit more if the round was short enough, and vice versa. 
\end{itemize}

\subsection{Stale Blocks and Forks}
A successfully mined block that is ultimately discarded from the longest chain is known as a `Stale block'. It happens because two or more miners can simultaneously solve the PoW puzzle for a specific round and create multiple blocks with different valid solutions. When this happens, a `fork' is created. A fork is a state in which there are conflicting opinions among network nodes regarding the status of the blockchain. In that case, only one block is added in the blockchain by fork resolution mechanism and other stale/orphan blocks are rejected. Transactions in these rejected blocks are sent back to the mempool (a cache used by every P2P node in a blockchain network) to await pickup in a subsequent block. When a fork occurs in Bitcoin
network by the miners, the longest chain rule is used to resolve it \cite{originalBitcoin}. The longest chain is the one that requires the most energy to construct. In Ethereum, fork resolution is based on the node with the heaviest sub-tree, which in short is called GHOST (Greedy Heaviest-Observed Sub-Tree) protocol \cite{sompolinsky2015secure}. However, recently Ethereum adopted a PoS-based consensus mechanism, which reduces the chance of fork \cite{buterin2020combining,buterin2017casper,ethereumPOS}.  

Forks also might occur for other reasons such as to implement a new feature, correct a security vulnerability, or settle a dispute within the community over the direction the blockchain system should take. In such situations, there are two types of fork: the hard fork and the soft fork. A hard fork is effectively a persistent deviation from the most recent version of a blockchain. This results in the blockchain being split into two separate networks that operate independently of one another since some nodes can no longer reach a consensus. Bitcoin Cash is an example of such a hard fork \cite{bitcoin2017Cash}. On the other hand, a soft fork is a backward compatible modification or upgrade of a blockchain. It does not cause the network to split off or produce a new version of the blockchain. Rather, it enables the network to smoothly switch over to the new rules while preserving compatibility with the previous ones. SegWit is an example of a soft fork that took place in 2017 for Bitcoin \cite{segwit2017activation}.

\begin{figure}[!h]
   \centering
  \includegraphics[width=1\linewidth]{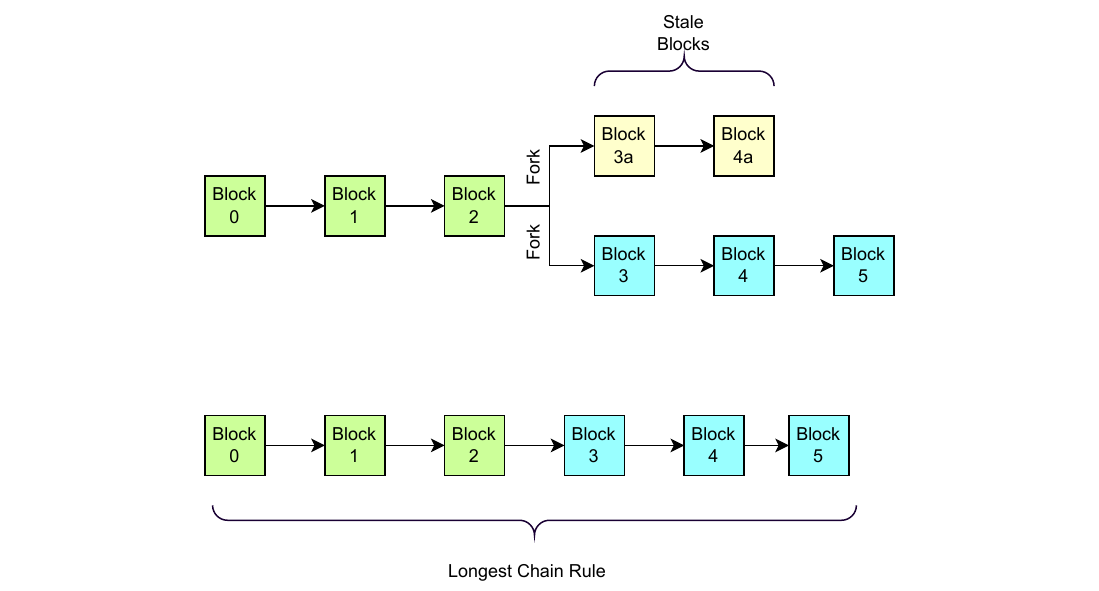}
  \caption{Fork and Longest Chain Rule}
   \label{fig:fork}
\end{figure}

Figure \ref{fig:fork} shows a hard fork, longest chain rule, and a scenario of stale blocks in a blockchain system.

\subsection{Double-Spending}
\label{double-spending}
Double-spending is a serious security risk in blockchain-based cryptocurrency systems. In contrast to a physical currency, which is tangible and cannot be replicated, digital currencies are inherently replicable. Double-spending takes advantage of this feature by attempting to spend the same digital currency several times, compromising the integrity of the system. We describe three attacks that are variants of the double spending attack in Section \ref{vector76_attack}, Section \ref{RaceAttack}, and Section \ref{FinneyAttack}. 

\section{Related Work}
\label{RelatedWorks}

A layer-based attack analysis is a systematic and efficient technique for critically analyzing a blockchain system. This method, which divides the system into smaller, more manageable components, makes it easier to identify and mitigate vulnerabilities by enabling a focused examination on certain components without being overwhelmed by the complexities of the entire system. Additionally, it makes it possible to introduce security measures that are layer-specific, putting the strongest protections where they are most needed and improving the system's overall resilience.

Other surveys have been conducted to gain insight into the attack surface of blockchain systems, which differ from the scope of our research. Guggenberger \textit{et al.} performed a review of the current literature on blockchain system attacks, ultimately identifying 87 relevant attacks. \cite{guggenberger2021structured}. The attacks were represented using the attack tree (AT) notation, as proposed by Mauw and Oostdijk (2006) \cite{mauw2006foundations}. However, a comprehensive analysis of each attack is absent, and the attacks are also not categorized in a layer-based approach.

Ferdous \textit{et al.} while introducing four distinct layers within a blockchain system, discussed some relevant attack vectors of consensus mechanisms, however, does not cover all layers \cite{ferdous2020blockchain}. 

In their study, Chen \textit{et al.} conducted a comprehensive analysis of the security aspects of the Ethereum system, encompassing vulnerabilities, assaults, and protection mechanisms \cite{chen2020survey}. The Ethereum platform and its associated layers, including the network, consensus, data, and application, were comprehensively addressed. In contrast, our work is not limited to one specific platform.

Li \textit{et al.} did a comprehensive analysis of the security vulnerabilities associated with widely used blockchain systems \cite{li2020survey}. The researchers conducted a blockchain security assessment by examining 20 distinct vulnerabilities, 6 attacks, and 5 corresponding defenses. Nevertheless, a layer-based categorization is missing in their work that we address. In addition, we analyze other relevant attacks.

Saad \textit{et al.} explored the attack surface of blockchains \cite{saad2020exploring} where 22 attacks and 33 defense mechanisms were covered. We take a different approach. We discuss attackers' incentives, specific vulnerabilities, stakes from attackers' perspective to launch such attack and categorize attacks using a layered approach. Furthermore, our analysis includes certain attacks that were absent in theirs.

Homoliak \textit{et al.} introduced the security reference architecture (SRA) as a framework for blockchains. Four layers were used in the study, including network, consensus, replicating state machine, and application. The ISO/IEC 15408 threat risk assessment standard was utilized for this purpose. Their study differs from ours in several ways. Firstly, we adopt the Application layer and the Meta-Application layer as suggested by \cite{ferdous2020blockchain}, which leads to different results and categorizations. Secondly, we illustrate how the components of different layers are impacted by attacks. Lastly, we utilize our own custom framework presented in Section \ref{taxonomy} to gain deeper insight into each attack.

Wen \textit{et al.} studied attacks and countermeasures using a six-layer blockchain model \cite{wen2021attacks}. The corresponding layers are as follows: Data, Network, Consensus, Incentive, Contract and Application. However, while organizing attacks, they combined the consensus and incentive layers and did not cover the application layer. In addition, they provided an overview and countermeasures for each attack. In contrast, we dive deep into each attack and explicitly analyze utilizing our framework. We also analyze inter-layer connections for attacks.

Zheng \textit{et al.} also proposed six layers: Data, Network, Consensus, Incentive, Contract, and Application \cite{zhang2019security}. Their research is primarily focused on the security and privacy features and techniques of blockchain, as well as the comparison of various consensus mechanisms. In discussing security and privacy properties, they briefly covered four major types of attacks and vulnerabilities. However, we developed a completely different strategy, diving deep into the categorization of layer-based attacks.

In addition, several studies have been conducted to categorize attacks on blockchain systems. Moubarak \textit{et al.} examined blockchain security, with a focus on Hyperledger, Ethereum, and Bitcoin \cite{moubarak2018blockchain}. They provided a summary of several challenges and attack scenarios and briefly reviewed possible mitigation techniques. Anita \textit{et al.} presented a taxonomy of security risks associated with blockchain technology, introducing 7 groups covering 17 attacks \cite{anita2019blockchain}. Chen \textit{et al.} proposed a blockchain attack classification system using three categories encompassing 11 attacks \cite{chen2022survey}. However, none of these studies used any layer-based categorization.

\section{Blockchain layers, property taxonomy and attacks}
\label{layerBasedAttackAnalysis}
In this section, we group attacks according to the layers they target. Many attacks affect multiple layers at the same time. In those situations, we highlight the attack in the most affected layer. Toward this aim, we present the layer-based approach in blockchain as introduced in \cite{ferdous2020blockchain} (Section \ref{subsec:lbc}). Then, we introduce the taxonomy of properties (Section \ref{taxonomy}) and present the list of attacks (Section \ref{subsec:attacks}).

\subsection{ Layer-based Approach on Blockchain}
\label{subsec:lbc}
\begin{figure}[!h]
\centering
\captionsetup{type=figure}
\begin{minipage}[b]{\columnwidth}
  \centering
  \includegraphics[width=0.7\linewidth]{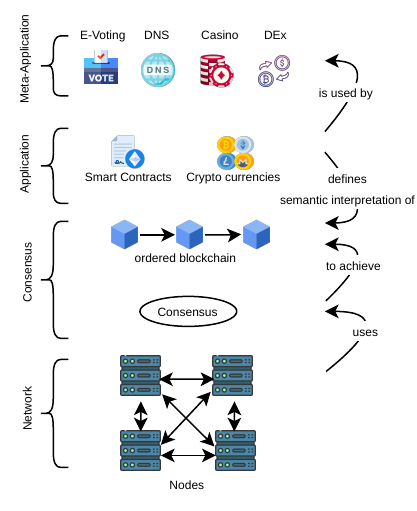}
\end{minipage}\hfill
\captionof{figure}{Blockchain Layers}
\label{fig:layers}
\end{figure}

In a blockchain system, various components exist, each responsible for performing one or more specific functions. Hence, it is essential to decompose the whole system into different layers to achieve modularity, scalability, security, and interoperability. David \textit{et al.} suggested four layers of blockchain: consensus, mining, propagation, and semantic \cite{david_layers}. However, they mostly focused on the blockchain system, ignored the P2P component altogether. Ferdous \textit{et al.} proposed another layer scheme with four layers: network, consensus, application, and meta-application \cite{ferdous2020blockchain} and this addressed the issue in the layer model of David \textit{et al.}. Our attack analysis is based on the scheme proposed by Ferdous \textit{et al.}. Here, we briefly discuss the layers and their responsibilities
\begin{itemize}
    \item \textbf{Meta-Application Layer:} The objective of a blockchain system's meta-application layer is to create an overlay atop the application layer so that other application domains can benefit from the logical interpretation of a blockchain system. 
    \item \textbf{Application Layer:} The logical interpretation of a blockchain system is determined by the application layer. A logical interpretation can be a cryptocurrency, a smart contract, a reward mechanism, etc.
    \item \textbf{Consensus Layer:} The distributed consensus is provided by the consensus layer. Various consensus algorithms such as PoW, PoS, etc. are the essential part of this layer. These algorithms are utilized to obtain the necessary consensus in the system.
    \item \textbf{Network Layer:} The network layer components are in charge of managing network capabilities, such as joining the peer-to-peer network, adhering to the networking protocol, informing newly joined nodes of the current status of the blockchain, propagating and receiving transactions and blocks, and other related tasks. Figure \ref{fig:layers} illustrates a high-level overview of blockchain layers.

\end{itemize}

\subsection{Taxonomy of properties}
\label{taxonomy}
We study each attack using the following properties:

\begin{enumerate}[label=\Roman*]
    \item \textbf{Motivation \& vulnerability:} This refers to the overview of attacker(s) intention, in regard to parts and properties of blockchain system that an attacker wants to take advantage of and  specific desirable states of a blockchain system for an attacker in certain attacks.
    \item \textbf{Attack strategy overview:} This property implies the generalized method of attack execution with simplified states.
    \item \textbf{Conditions \& outcomes:} The term reflects system condition(s) or fact(s) that affects how the attack turns out \& which parts of the system strongly impacted by an attack(s). This also elaborates the risk factors regarding an attacker's motive \& consequences of a successful attack(s) in a system.
    \item \textbf{Enhancements:} This property explains the connection to other attacks or leveraging situations where the current attack may lead to another.
    \item \textbf{Plausibility \& prevention measures:} This indicates the availability of a solution to a specific assault and provides a view of the practical difficulty of executing the attack.  
\end{enumerate}

\subsection{Analyzed attacks}
\label{subsec:attacks}
The list of attacks analyzed in this article is presented in Table \ref{tab:attack_list}. In order to compile this list of attacks, we conducted a comprehensive literature review in several sources including the arXiv pre-print server, journals and conference proceedings. It was based on the frequency and consequences of such attacks as reported in earlier studies in those sources.By incorporating multiple credible sources, we ensured that the list encompassed the latest and most prevalent attacks in the field. This systematized approach provides an excellent basis for the examination carried out in the present work.
\begin{table}[h]
\caption{Surveyed Attacks}
\label{tab:attack_list}
\begin{tabular}{c|c|c|c}
\hline
\rowcolor[HTML]{E9E9E9} 
\textbf{Network Layer} & \textbf{Consensus Layer} & \textbf{Application Layer} & \textbf{\begin{tabular}[c]{@{}c@{}}Meta-Application\\ Layer\end{tabular}} \\ \hline
Balance Attack         & Punitive Forking Attack  & Replay Attack              & \multicolumn{1}{l}{Front Running Attack}                                 \\ \hline 
\rowcolor[HTML]{EEEEEE} 
Sybil Attack         & Block Withholding Attack  & Short Address &    Block Stuffing Attack                      \\ \hline
Eclipse Attack       & Fork After Withholding    &               &                          \\ \hline
\rowcolor[HTML]{EEEEEE} 
BGP Hijacking attack & Vector 76                 &               & \cellcolor[HTML]{E2DFDF} \\ \hline
Pool Hopping attack  & Selfish Mining Attack     &               &                          \\ \hline
\rowcolor[HTML]{EEEEEE} 
                     & Race Attack               &               &                          \\ \hline
                     & Finney Attack             &               &                          \\ \hline
\rowcolor[HTML]{EEEEEE} 
                     & Long Range (LR) : Simple  &               &                          \\ \hline
                     & LR : Posterior Corruption &               &                          \\ \hline
\rowcolor[HTML]{EEEEEE} 
                     & LR : Stake Bleeding       &               &                          \\ \hline
                     & P+Epsilon Attack          &               &                          \\ \hline
\rowcolor[HTML]{EEEEEE} 
                     & Feather Forking Attack    &               &                          \\ \hline
                     & Bribery Attack            &               &                          \\ \hline
\rowcolor[HTML]{EEEEEE} 
                     & Consensus Delay Attack      &               &                        \\ \hline
\end{tabular}
\end{table}

\section{Network Layer Attacks}
\label{sec:netLayerAttack}
Network layer attacks typically interrupt node communications and target processes that allow nodes to communicate and agree on data, exploit vulnerabilities in node detection, transaction/block propagation, and communication protocols, and disrupt the functions of the network layer, such as data transfer and synchronization. Communication channels underpin blockchain networks; therefore, network layer attacks target them. These channel disruptions have a direct influence on network performance. The attacks analyzed under the network layers are discussed next.

\subsection{Balance Attack}
\label{BalanceAttack}
The Balance attack was first identified by Christopher Natoli and Vincent Gramoli in 2017, which targets the blockchain fork mechanism\cite{8023156BALANCE00}. The main target of the attack is Ethereum; however, Bitcoin is also vulnerable to the same strategy. It is to be noted that this attack is applicable to both the network and consensus layers. Next, we analyze this attack using the selected properties.

\vspace{2mm}
\noindent \textbf{Motivation \& vulnerability:} In this attack, the attacker leverages Ethereum's Ghost Protocol \cite{sompolinsky2015secure} or, in the context of Bitcoin, the longest chain rule. The objective is to hinder block propagation across the network and achieve double-spending \cite{app9091788BALANCE01}, by manipulating the selection of branches.

\begin{figure}[h]  
    \centering
    \includegraphics[width=0.8\textwidth]{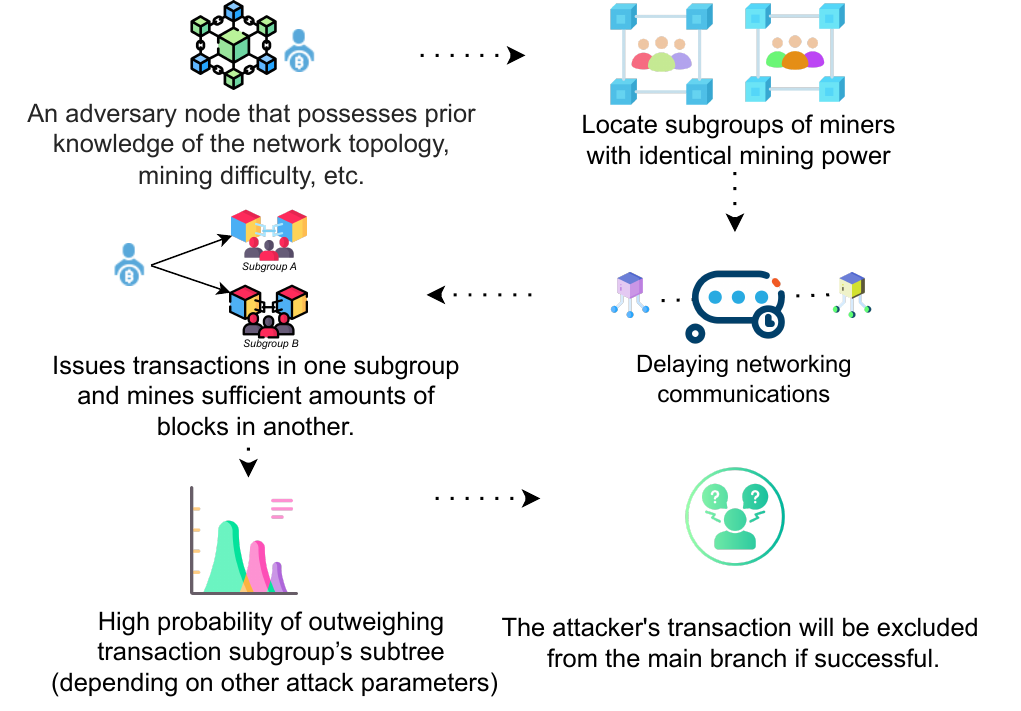} 
    \caption{Overview of Balance Attack}
    \label{fig:balance_attack}
\end{figure}

\vspace{2mm}
\noindent \textbf{Attack strategy overview:} The attack strategy overview is presented in Figure \ref{fig:balance_attack}.

\noindent \textbf{Conditions \& outcomes:} The attack requires key components such as knowledge of miners' logical or physical communication graphs, computational capacity for mining, and the current difficulty level \cite{8023156BALANCE00}. A Successful attack will damage honest miners, making mining efforts ineffective for one subgroup and potentially harming intended payment recipients.

\vspace{2mm}
\noindent \textbf{Enhancements:} Regardless of Balance attack success, an attacker may develop a consensus delay attack situation by trying to establish a reasonable delay between nodes, as explained later in our work in Section \ref{ConsensusDelayAttack}.

\vspace{2mm}
\noindent \textbf{Plausibility \& prevention measures:} This attack is feasible with limited mining power and dynamic data from multiple sources, including blockchain communication architecture, such as propagation latency, difficulty, and linked nodes. So far, No mitigating methods for forkable blockchains have been presented for this attack \cite{8023156BALANCE00}.

\subsection{Sybil Attack}
\label{SybilAttack}
First put forward by John R. Douceur in his 2002 paper "The Sybil Attack" \cite{10.1007/3-540-45748-8_24SybilAttack01}, Sybil attack is an attempt to dominate a network by leveraging many aliases. Successful Attackers can host many nodes and outvote reliable users \cite{BinanceAcademy_2020,ImpervaSybilAttack}. Sybil attacks are a milder form of an attack known as a \textit{51\% attack}. In a 51\% attack, if malicious miners happen to control more than 50\% of the network hash rate, they manipulate the blockchain protocol rules. 

\vspace{2mm}
\noindent \textbf{Motivation \& vulnerability:} With the ultimate motive to corrupt the peer-to-peer network, the attacker aims to create and maintain large numbers of digital identities ( assisted by multiple devices, virtual machines, IP addresses, and botnets \cite{app9091788BALANCE01,swathi2019preventingSybilAttack00} ) and use these identities to capture new nodes by exploiting node limitations in state synchronization \cite{10.5555/2695500SybilAttack04}.

\begin{figure}[!htbp]  
    \centering
    \includegraphics[width=0.85\textwidth,  keepaspectratio]{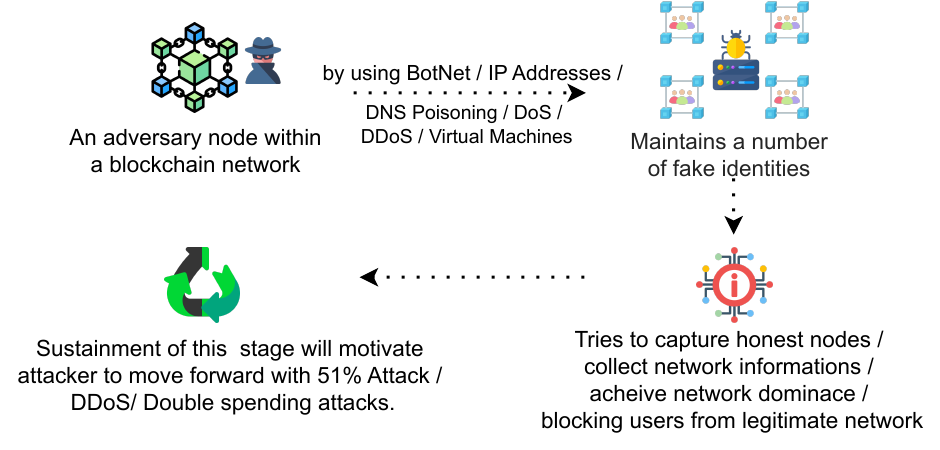}
    \caption{Overview of Sybil Attack}
    \label{fig:sybil_attack}
\end{figure}

\vspace{2mm}
\noindent \textbf{Attack strategy overview :} The attack strategy overview is presented in Figure \ref{fig:sybil_attack}.

\noindent \textbf{Conditions \& outcomes :} The number of Sybil nodes controlled by the attacker and cooperation among them determine the attack severity. The attacker can influence consensus at specific lengths and lower the block propagation performance if a fake network is developed \cite{ImpervaSybilAttack}. Due to the attacker's actions, the majority bar is raised, wasting more processing power \cite{swathi2019preventingSybilAttack00}.

\vspace{2mm}
\noindent \textbf{Enhancements:} A successful attack can lead to consensus delay, DoS and DDoS attacks \cite{swathi2019preventingSybilAttack00}, or 51\% attacks \cite{ImpervaSybilAttack}. Also, it  raises the chance of double-spending attacks like Finney Attack and Race Attack, outlined later in our work.

\vspace{2mm}
\noindent \textbf{Plausibility \& prevention measures:} In ‘The Sybil Attack,’ John R. Douceur said that Sybil attacks are always possible without any logically centralized authority, such as in the public blockchain \cite{10.1007/3-540-45748-8_24SybilAttack01}. Public blockchains have been attacked by Sybil attacks in the past \cite{SolSybilAttack03}.
In addition, Bitcoin created the Bloom Filter to protect and minimize privacy threats for lightweight nodes, or Simplified Payment Verification (SPV) nodes \cite{10.5555/2695500SybilAttack04}. Different studies show that this Bloom Filter does not fully protect a node from synchronizing with a Sybil node or agent \cite{9333950BloomFilter}.

Additionally, a wallet-generated address method has been proposed to identify Sybil attacks on public blockchains \cite{swathi2019preventingSybilAttack00}. Direct and indirect identity validation methods have been proposed to counter Sybil attacks \cite{10.1007/3-540-45748-8_24SybilAttack01}. Several strategies, including registration-based methods, third-party mixing protocols, neighborhood similarity, network clustering, and position verification, have been proposed to prevent Sybil attacks \cite{swathi2019preventingSybilAttack00}. Some of these methods are also applied to specific networks but not in major public blockchains. In addition, application-specific defenses are useful to defend against systemic Sybil attacks \cite{ImpervaSybilAttack}.

\subsection{Eclipse Attack}
\label{EclipseAttack}
An Eclipse attack happens when an attacker segregates a particular user or node within a peer-to-peer (P2P) network. The idea of an Eclipse attack on blockchain was first discussed by Heilman \textit{et al} \cite{heilman2015Eclipse01}. Although the steps of this attack are somewhat similar to the Sybil attack, the motivations and objectives are distinct \cite{eclipseEclipse03}.

\vspace{2mm}
\noindent \textbf{Motivation \& vulnerability:} Eclipse attacks exploit blockchain's peer-to-peer protocol limitations (i.e. - outgoing and incoming connections rule, persistent network information \cite{heilman2015Eclipse01}, node state synchronization \cite{karl2016ethereumEclipse02}) to isolate a node from the network. Next, the attacker entirely controls the victim's information access to restrict their blockchain view or co-opt their computing power for additional attacks \cite{marcus2018lowEclipse04}.

\begin{figure}[!htbp]  
    \centering
    \includegraphics[width=0.85\textwidth, keepaspectratio]{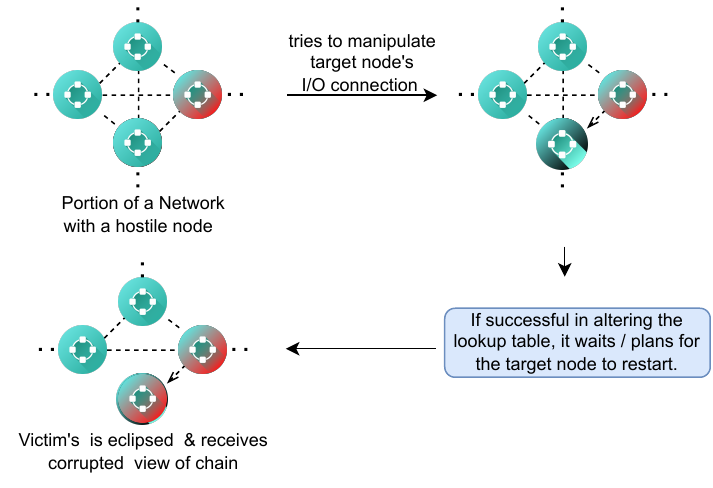}
    \caption{Overview of Eclipse Attack}
    \label{fig:eclipse_attack}
\end{figure}

\vspace{2mm}
\noindent \textbf{Attack strategy overview:} The attack strategy overview is presented in Figure \ref{fig:eclipse_attack}.

\noindent \textbf{Conditions \& outcomes:} Depending on the blockchain, its client's type, and the attacker's choice, there are many determining factors in eclipsing a node. Most common scenarios require some pre-computation \cite{karl2016ethereumEclipse02}, lookup table, and outgoing connection alteration, along with ensuring victim node's restart \cite{heilman2015Eclipse01}. Successful eclipsing disconnects the victim from the genuine blockchain state, which the attacker exploits.

\vspace{2mm}
\noindent \textbf{Enhancements:} If the victim is successfully eclipsed, it gives the attacker advantages in block racing, enables transaction hiding and selfish mining scenarios, 0/N confirmation double-spending and stake bleeding attack. Also it facilitates the wastage of the victim's mining power \cite{heilman2015Eclipse01,karl2016ethereumEclipse02,marcus2018lowEclipse04,eclipseEclipse03,zhang2019eclipse}. 

\vspace{2mm}
\noindent \textbf{Plausibility \& prevention measures:} Without a centralized authority, the attack is feasible, although resource utilization may vary \cite{marcus2018lowEclipse04}. To stand against eclipse attacks in Etherium , solutions like eliminating the reboot exploitation window, ensuring constant seeding, and limiting incoming TCP connections have been implemented \cite{marcus2018lowEclipse04}. Bitcoin implemented countermeasures include deterministic random eviction, random selection, increasing bucket numbers, feeler connections, and test before evict \cite{Heilman_2015,heilman2015Eclipse01}. 

\subsection{BGP hijacking Attack }
\label{BGPHijackingAttack}
This attack exploits the Border Gateway Protocol (BGP) routing protocol. Established in 1989, BGP is used to determine routing decisions among autonomous systems (ASes) on the Internet. BGP Hijacking involves manipulating internet routing tables using the protocol and illegitimately obtaining clusters of IP addresses. Through the BGP hijack attack, attackers can impersonate the IP address of their targets \cite{WhyBGPhijackingBGP2}. This attack poses a threat to blockchain by allowing hostile actors to redirect mining pools, resulting in income loss. In 2014, around \$83,000 (USD) was stolen by the BGP hijacking attack \cite{stewart_2014BGP0}. On 17 August 2022, an assailant carried out a BGP hijack on Celer Bridge, a cryptocurrency service, leading to the loss of \$235,000 worth of bitcoin \cite{Madory_2022BGP8}.

\vspace{2mm}
\noindent \textbf{Motivation \& vulnerability:} The attacker utilizes BGP route redirection to hijack cryptocurrency mining operations without payment, capturing honest miners in a malicious pool and maintaining the false pool for each honest miner in brief intervals. To keep the activity undetected \cite{stewart_2014BGP0}, the attacker can partition the network or intercept a portion of connections to induce delays in blockchain traffic through BGP hijacks \cite{apostolaki2017hijackingBGP1}.

\begin{figure*}[!h]  
    \centering
    \includegraphics[width=0.9\textwidth]{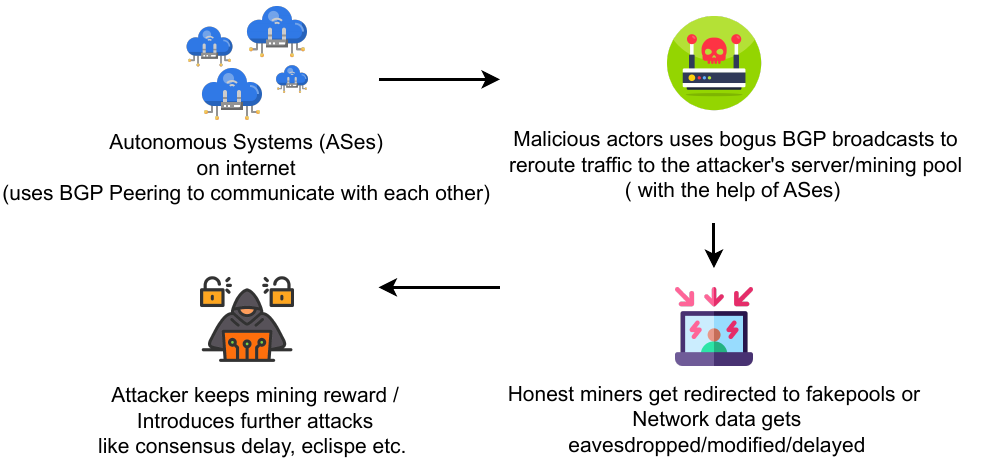}  
    \caption{Overview of BGP hijacking Attack}
\label{fig:bgp_hijacking_attack}
\end{figure*}

\vspace{2mm}
\noindent \textbf{Attack strategy overview:} The attack strategy overview is presented in Figure \ref{fig:bgp_hijacking_attack}.

\noindent \textbf{Conditions \& outcomes:} Access to Internet Routing Table is crucial, and efficiently corrupting it indicates susceptibility to the attack. It will leave miners in a rewardless state \cite{apostolaki2017hijackingBGP1}. Also, the attacker needs to ensure the bogus BGP announcements are unfiltered by the upstream network. Moreover, a BGP attack may lead to unexpected forks in the chain \cite{Madory_2022BGP8}.

\vspace{2mm}
\noindent \textbf{Enhancements:} A successful attack can isolate a node, creating a barrier in communication that increases vulnerability in multiple domains such as node eclipse, 0-confirmation double-spends, selfish mining, and consensus delay \cite{apostolaki2017hijackingBGP1}.

\vspace{2mm}
\noindent \textbf{Plausibility \& prevention measures:} One proposed solution is RouteChain, which utilizes a blockchain-based routing system to address BGP Hijacking \cite{saad2022routechainBGP3}. Similar solutions like as BlockJack, which is developed using Hyperledger Fabric and Quagga, are also recommended \cite{sentana2021blockjackBGP4}. Lukas \textit{et al}. proposed an alternative method to improve the security in the BGP Protocol by overseeing AS border routers \cite{mastilak2020enhancingBGP6}. For long-term defenses, employing separate control and data channels, alongside UDP heartbeats and encrypted communications, serve as effective solutions \cite{apostolaki2017hijackingBGP1}.

\subsection{Pool Hopping Attack }
\label{PoolHoppingAttack}
A mining Pool is formed by miners to speed up the new block mining process and receive rewards with accumulated effort and less difficulty. Depending on where the pool is right now, predicted profits, volatility, and maturity time will change. The practice of mining just when the payoff is high and the difficulty is low and leaving when the opposite is true is known as pool hopping \cite{singh2019smartPoolHopping1}. Thus, hoppable pool mining is unfavorable, as more miners hop, the pool becomes increasingly unstable. Everyone mining alone or at a hopping-proof or very hopping-resistant pool is the only sustainable approach \cite{rosenfeld2011analysis}.

\vspace{2mm}
\noindent \textbf{Motivation \& vulnerability:} Various mining pools employ different reward methods as explained in Section \ref{subsec:miningPool}. A logical miner has a choice to join a larger reward pool to maximize the monetary outcome of their mining efforts. The exploitation of block reward distribution can be achieved by the analysis of mining pool behavior and the selection of an effective mining method \cite{tyab027PoolHopping4}. Therefore, the assailant has the ability to develop a steady level of revenue with pool hopping.

\begin{figure}[!bp]
    \centering
    \begin{minipage}[b]{0.48\textwidth}
        \includegraphics[width=0.95\textwidth]{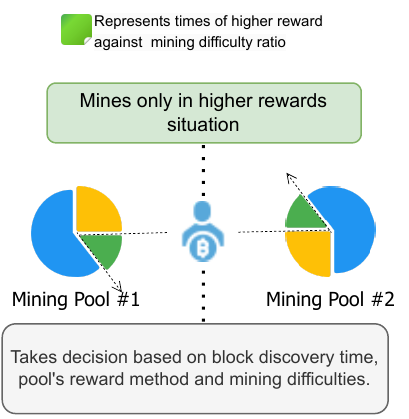}
        \captionof{figure}{Overview of Pool Hopping Attack}
        \label{fig:pool_hopping_attack}
    \end{minipage}
    \begin{minipage}[b]{0.50\textwidth}
         \hspace{-0.57cm}
        \includegraphics[width=1.1\textwidth]{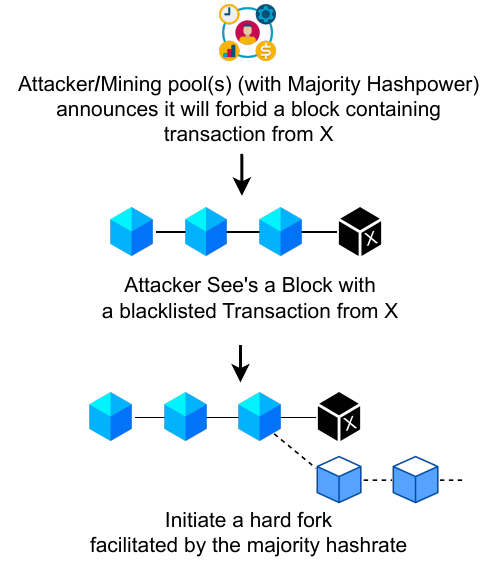}
        \captionof{figure}{Overview of Punitive Forking}
        \label{fig:punitive_forking_attack}
    \end{minipage}
\end{figure}

\vspace{2mm}
\noindent \textbf{Attack strategy overview:} The attack strategy overview is presented in Figure \ref{fig:pool_hopping_attack}.

\vspace{2mm}
\noindent \textbf{Conditions \& outcomes:} For attackers to execute this attack successfully, they need to time the pool switch effectively, and keep track of longer mining rounds \cite{8473692PoolHopping2}, network delays in pool joining, and the current hashrate of pool \cite{rosenfeld2011analysis,8473692PoolHopping2}. If executed properly, it will impact the pool's honest miners, the pool's share and fluctuations in total hashrate of the pool. Overall, an unsustainable mining pool with increased block generation time and and an increased chance for an attacker to secure roughly 28\% more rewards, depending on the hashrate ratio between between hoppers and honest miners \cite{rosenfeld2011analysis}.

\vspace{2mm}
\noindent \textbf{Enhancements:} This attack does not lead to another attack.

\vspace{2mm}
\noindent \textbf{Plausibility \& prevention measures:} There have been previously created technologies, such as Bithopper, for pool hopping that have been successful \cite{PoolHopping3}. Proposed solution to mitigation of Pool hopping includes mining resource allocation mechanism (developed with python and cobyla) \cite{tyab027PoolHopping4}. A pool hopping prevention strategy using a smart contract is also proposed which aims to detect and prevent the attacks using miner certificates and by introducing risk factors in pool switching during incomplete block mining \cite{singh2019smartPoolHopping1}. In addition to these, a novel mining pool design is also proposed , based on zero-determinant theory and iterated prisoner’s dilemma (IPD) game which is fee-free and introduces fairness as long as mutual cooperation exist between miners and the pool \cite{9186838PoolHopping5}.

\section{Consensus Layer Attacks}
\label{sec:conLayerAttack}
Blockchain systems use a consensus method to ensure that all participants agree on the current state of the ledger. This process serves as the foundation for trust and security in any blockchain system. Malicious actors can disrupt the consensus method used by participating nodes to validate malicious transactions and manipulate the blockchain state. These attacks aim to disrupt, corrupt, or misuse the consensus process by taking advantage of weaknesses in the consensus algorithm of the blockchain system, potentially causing catastrophic effects for the blockchain. The attacks in the consensus layer are discussed next.

\subsection{Punitive Forking Attack }
\label{PunitiveForkingAttack}
Punitive forking refers to the act of targeting a specific entity in order to enforce laws or ban its transactions. The attack targets the consensus layer and compromises the integrity of the chain, potentially leading to the centralization of hashpower \cite{conti2018surveyPunitive_Forking1}.

\vspace{2mm}
\noindent \textbf{Motivation \& vulnerability:} This attack can be carried out in a variety of ways by exploiting hashpower advantage. It is also viable to announce an address blacklisted and enforce it for pools owned by the attacker if they own more than 51\% of the network hashrate \cite{howtoDestroyPunitive_Forking2,Punitive_Forking0}.

\vspace{2mm}
\noindent \textbf{Attack strategy overview:} The attack strategy overview is presented in Figure \ref{fig:punitive_forking_attack}.

\vspace{2mm}
\noindent \textbf{Conditions \& outcomes:} It is required by the attackers to possess more than 51\% of the network's hashpower in order to convince miners to suspend all transactions from a particular address. Hard forks and restricted addresses are the outcomes of successful assaults \cite{Iosr_jeeePunitive_Forking3,Hypothetical_at_Feather_Forking2}. It also has an effect on the reliability of a lesser part (<=49\%) of the consensus as well.

\vspace{2mm}
\noindent \textbf{Enhancements:} This attack does not provide any leverage for subsequent attacks.

\vspace{2mm}
\noindent \textbf{Plausibility \& prevention measures:} Despite blockchain's decentralized nature, mining pools can, in the worst-case scenario, consolidate hashpowers that could be used for this assault, making any solution proposal impractical at this time. Currently, achieving the goal of hashpower domination on the network requires the accumulation of hashpower from a just 2 or 3 mining pools \cite{miningPoolsSummaryBest_2024}. Thus, the prevention of punitive forking remains an open challenge \cite{conti2018surveyPunitive_Forking1}.

\subsection{Block Withholding (BWH) Attack}
\label{BlockWithHelding}
Rosenfeld et al. introduced the Block Withholding (BWH) attack \cite{rosenfeld2011analysis}. In this attack, the attacker or a group of attackers prevents the mining pool from receiving a legitimate block to affect the pool manager or the other honest pool miners. 

\vspace{2mm}
\noindent \textbf{Motivation \& vulnerability:} The motivation of this attack is to increase profits from mining or to harm the pool manager. The attack impairs the mining pool's reward-sharing scheme. This attack can be carried out by a single dishonest miner or a group of dishonest miners with a significant quantity of mining power. Also, this attack can be launched by a pool to another pool \cite{eyal2015miner}. Most of the mining pools are open. Miners can join in these pools through a public Internet interface. However, these open pools are susceptible to traditional block withholding attacks.

\begin{figure}[!h]
    \begin{minipage}[b]{0.48\textwidth}
        \centering
        \includegraphics[width=\textwidth,keepaspectratio]{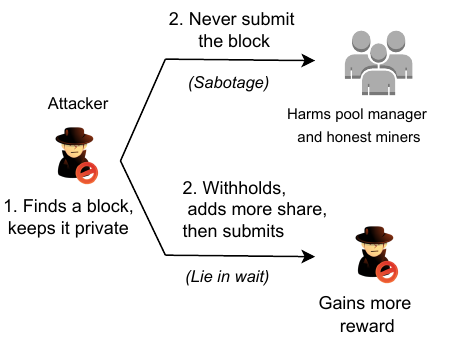}
        \captionof{figure}{Overview of Block Withholding Attack}
        \label{fig:block_withholding}
    \end{minipage}
    \hfill    
     \begin{minipage}[b]{0.50\textwidth}
        \centering
        \includegraphics[width=\textwidth,keepaspectratio]{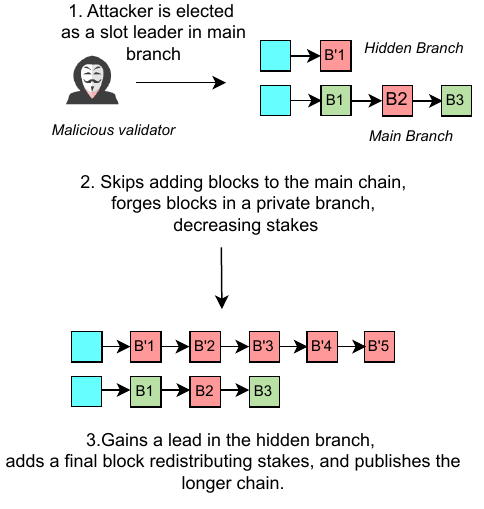}
        \captionof{figure}{Overview of Long Range Attack (Stake Bleeding)}
        \label{fig:long_range_stake_bleeding}
    \end{minipage}
    \hfill
\end{figure}

\vspace{2mm}
\noindent \textbf{Attack strategy overview:} Rosenfeld talks about two different variants of this attack:  Sabotage and Lie in Wait \cite{rosenfeld2011analysis}. In `Sabotage', if a dishonest miner finds a `share', he never submits it to the pool manager. In `Lie In Wait', if the attacker solves a block, he withholds it and keeps adding his share to the pool. After a certain amount of time, he submits the block. Thus he receives more profit. The attack strategy overview is presented in Figure \ref{fig:block_withholding}.

\noindent \textbf{Conditions \& outcomes:} This attack has a significant relation with the attacker's hash power as he needs to add more shares or solve a block. This attack, in any form, negatively impacts both the pool operator and the honest miners who operate within the pool. If the PPS reward sharing system is applied in the pool as described in Section \ref{subsec:miningPool}, the manager has to bear the loss. The `Lie in Wait' variant can be useful for the attacker to gain monetary value. However, Curtois \textit{et al.} argued that the `Subotage' variant can also bring monetary value to the attacker \cite{courtois2014subversive}.

\vspace{2mm}
\noindent \textbf{Enhancements:} This attack can lead to Race attack described in Section \ref{RaceAttack} \cite{eyal2015miner}.

\vspace{2mm}
\noindent \textbf{Plausibility \& prevention measures:} In 2014, the `Eligius' mining pool experienced a 300 BTC loss \cite{wizEligius}, demonstrating the attack's feasibility. A two-step PoW mechanism called `Oblivious Share' is proposed to prevent this attack, but it wastes miners’ computing resources \cite{rosenfeld2011analysis}. To avoid this attack, open mining pools should only include known and trusted individuals \cite{courtois2014subversive}. 

\subsection{Fork After Withholding (FAW)}
\label{ForkAfterWithhelding}
This is another variant of the Block Withholding attack, but with a different goal and result. Kwon \textit{et al.} first proposed this strategy and came to the conclusion that this attack is always advantageous to the attacker \cite{kwon2017selfish}. 

\vspace{2mm}
\noindent \textbf{Motivation \& vulnerability:} The motivation of this attack is to gain profit by creating an intentional fork in the blockchain. This attack can be carried out by a single miner or a pool against another pool. Similar to the BWH attack, the open pools are vulnerable to the FAW attack.

\begin{figure}[b]
    \begin{minipage}[b]{1\textwidth}
        \centering
        \includegraphics[width=\textwidth,keepaspectratio]{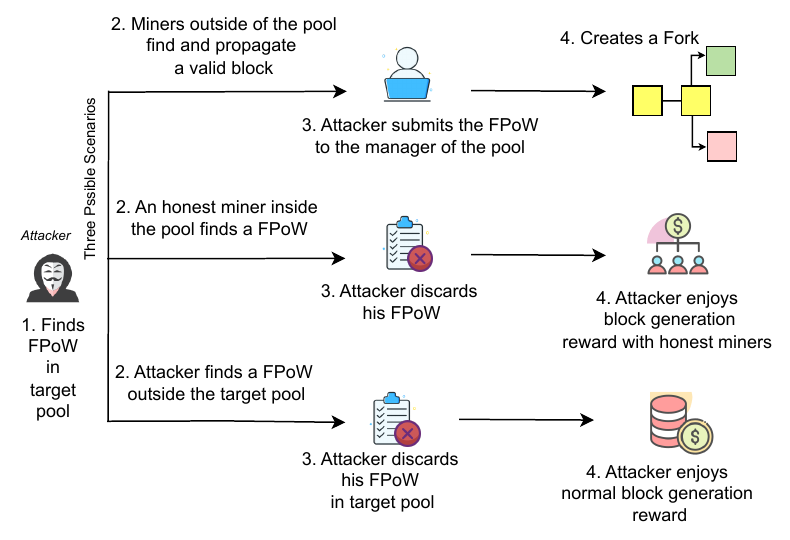}
        \captionof{figure}{Overview of Fork After Withholding Attack}
        \label{fig:fork_after_withholding}
    \end{minipage}
\end{figure}

\vspace{2mm}
\noindent \textbf{Attack strategy overview:} In this attack, the attacker splits his mining power into two different fractions. One part is invested into the target mining pool, and the other part is utilized for legitimate solo mining. The rest of the attack strategy is briefly presented in Figure \ref{fig:fork_after_withholding}.

Similarly, an attacker can simultaneously target numerous pools and distribute his mining power among them. This approach creates a fork with n+1 branches by focusing on n pools. Furthermore, an attack can be launched by two pools against one another.

\vspace{2mm}
\noindent \textbf{Conditions \& outcomes:} The attacker needs find an FPoW. Mining power plays a vital role here. He needs to split his mining power optimally. Therefore, he must be aware of the computational power of the target pool and the likelihood that his FPoW will be chosen for the main chain \cite{kwon2017selfish}. If the attacker possesses Sybil nodes, it will be beneficial for him to notice external block propagation faster. A successful attack may lead to a fork in the blockchain or increase the attacker’s rewards. The FAW attack can significantly increase an attacker's rewards by up to four times in comparison to a BWH attacker \cite{kwon2017selfish}.

\vspace{2mm}
\noindent \textbf{Enhancements:} This attack can lead to Race attack as it is similar to BWH attack.

\vspace{2mm}
\noindent \textbf{Plausibility \& prevention measures:} This attack is feasible and can be launched against Ethereum, Dogecoin, Permacoin and Litecoin as well as Bitcoin blockchain systems \cite{kwon2017selfish}. Several methods have been proposed to prevent this attack, like `Oblivious Share' \cite{rosenfeld2011analysis} and `Two Phase Proof of Work' \cite{eyal2014disincentivize}. In addition, Kwon \textit{ et al. } \cite{kwon2017selfish} propose a reward-sharing mechanism which decreases the risk of this attack but causes high reward variance. 

\subsection{Vector76 Attack}
\label{vector76_attack}
This attack was initially suggested by a user named vector76 in a bitcoin forum \cite{vector76}. It combines the elements of the Race attack and the Finney attack (discussed later). Here, the attacker pre-mines a block and then, attempts to broadcast and add the pre-mined block to the main chain. This is also known as a `1-confirmation' attack.

\vspace{2mm}
\noindent \textbf{Motivation \& vulnerability:} The cryptocurrency exchanges that allow for deposits and withdrawals of funds are the main targets of this attack. A cryptocurrency exchange is a digital platform that allows traders to exchange cryptocurrencies for various assets, including fiat currencies or other cryptocurrencies. If the exchange service allows an incoming connection, the attacker gets a chance to launch this attack. 

\begin{figure}[!h]
    \begin{minipage}[b]{1\textwidth}
        \centering
        \includegraphics[width=\textwidth,keepaspectratio]{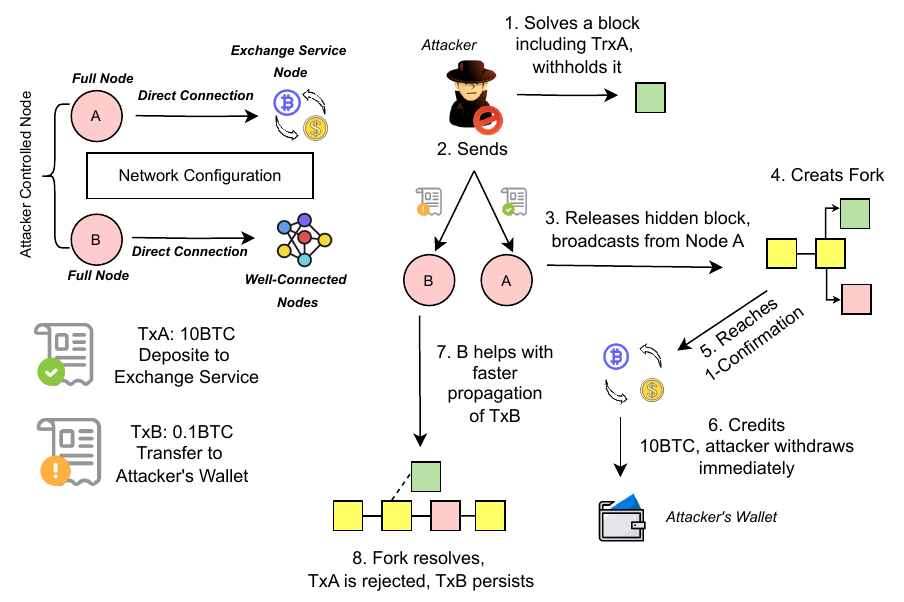}
        \captionof{figure}{Overview of Vector 76 Attack}
        \label{fig:vector76_attack}
    \end{minipage}
    \hfill
\end{figure}

\vspace{2mm}
\noindent \textbf{Attack strategy overview:} The attack strategy overview is presented in Figure \ref{fig:vector76_attack}.

\vspace{2mm}
\noindent \textbf{Conditions \& outcomes:} The attacker must pre-mine a block. The targeted exchange service must allow other nodes to discover and directly connect to their nodes. A successful attacker keeps the withdrawn money along with the later transaction as shown in Figure \ref{fig:vector76_attack}. In the event of a failed attack when the pre-mined block is not added to the main chain, the attacker will still have a deposit in the exchange. If the pre-mined block remains in the main chain, the block generation reward also goes to the attacker.

\vspace{2mm}
\noindent \textbf{Enhancements:} This attack does not provide any leverage for subsequent attacks.

\vspace{2mm}
\noindent \textbf{Plausibility \& prevention measures:} There have been no occurrences of this attack thus far. An effective mitigation strategy involves employing a waiting period for transactions until numerous confirmations are received. Additionally, exchange services should refrain from accepting incoming connections. 

\subsection{Selfish Mining Attack}
\label{SelfishMiningAttack}
Eyal \textit{et al.} first modeled this attack on the Bitcoin system \cite{eyal2014majority}. In this attack, the attackers purposefully keep their blocks secret to maximize the rewards. Instead of adding their discovered blocks to the public blockchain, the attackers build a separate, private version of the blockchain and continue to add new blocks to it. Meanwhile, honest miners, unaware of the existence of the private chain, continue to mine on the public version of the chain, which is actually lagging behind. The race between public and private versions of the same chain ends when the attacker publishes the longest private chain. If the network is built upon the longest chain rule, the network adopts the attackers' chain, causing the honest miners' work to be invalidated.

\vspace{2mm}
\noindent \textbf{Motivation \& vulnerability:} Attackers want to enhance their rewards and control the network by withholding mined blocks and surreptitiously mining on top of them. This approach uses the longest chain rule in the proof-of-work consensus mechanism to fork the blockchain.

\begin{figure*}[!h]  
    \centering
    \includegraphics[width=1.0\textwidth]{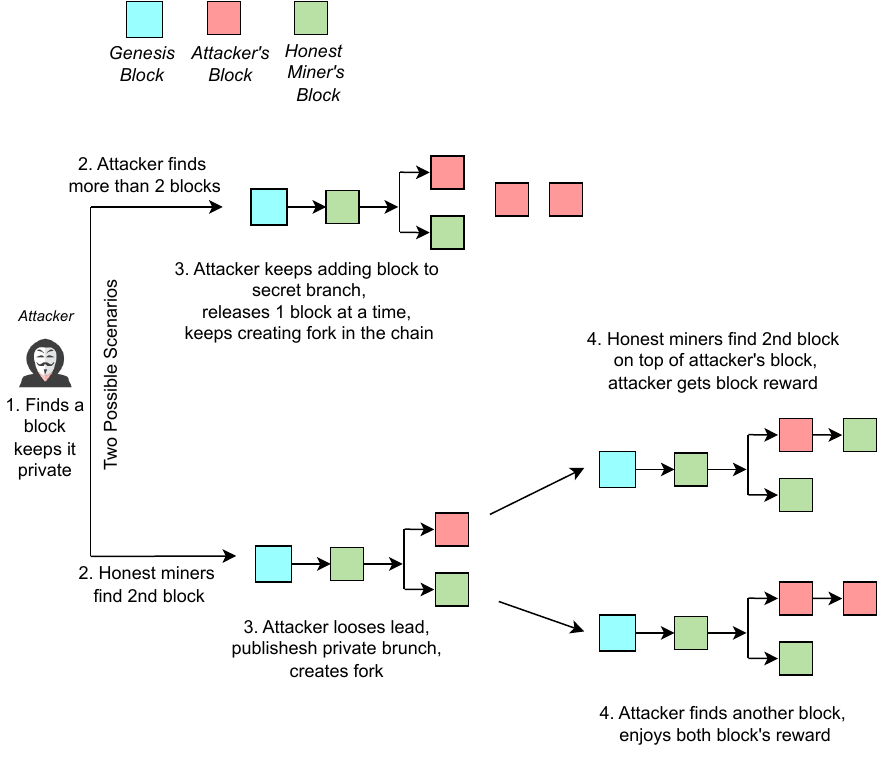}  
    \caption{Overview of Selfish Mining Attack}
\label{fig:selfish_mining}
\end{figure*}

\vspace{2mm}
\noindent \textbf{Attack strategy overview:} The attack strategy overview is presented in Figure \ref{fig:selfish_mining}. This attack can also take place in a mining pool. The malicious pool can act as a single agent and repeat the event \cite{eyal2014majority}.

\vspace{2mm}
\noindent \textbf{Conditions \& outcomes:} The attacker must initiate the process by discovering a valid block. The mining power of the attacker is crucial in this context. According to \cite{kwon2017selfish}, 9\% hash power is required. In addition, to achieve a successful fork, the attacker must transmit the block more rapidly over the network. The consequences of such an attack are significant. It interferes with the consensus protocol and compromises the integrity of the system. Miners expend their resources on unproductive blocks. Furthermore, the inclusion of all transactions in the block of the honest miner is likewise met with rejection, thus creating opportunities for various forms of other attacks.

\vspace{2mm}
\noindent \textbf{Enhancements:} A successful attack can lead to a double-spending and fork after withholding attack.

\vspace{2mm}
\noindent \textbf{Plausibility \& prevention measures:}
This attack is considered to be impractical \cite{kwon2017selfish}. To counter this attack, a timestamp-based solution is proposed by Solat \textit{et al.} \cite{solat2016zeroblock}. Saad \textit{et al.} proposed the concept of `truth state' for blocks and included the expected confirmation height parameter in transaction data structures \cite{saad2019countering}.

\subsection{Race Attack}
\label{RaceAttack}
Karame \textit{et al.} proposed and modeled this attack \cite{karame2012two}. This is an example of a `double-spending' attack, where an attacker tries to use the same currency for two separate transactions.

\vspace{2mm}
\noindent \textbf{Motivation \& vulnerability:} The `Fast Payment System' of Bitcoin is exploited by this attack. On average, a new block generation in Bitcoin network takes approximately ten minutes \cite{karame2012two}. It is clear that while taking Bitcoin payments, vendors and merchants such as supermarkets, take-away stores, vending machines, etc. cannot rely on transaction completion (i.e. new blocks being added and having enough confirmation). Therefore, as soon as the network transmits a transaction containing the required amount of BTCs from the client to one of its addresses, the merchant can accept bitcoin payments with no confirmations for low-cost transactions \cite{bitcoin_faq}. The attacker can use this loophole to their advantage and perform this attack.

\begin{figure}[b]
    \begin{minipage}[b]{0.50\textwidth}
        \centering
        \includegraphics[width=\textwidth,keepaspectratio]{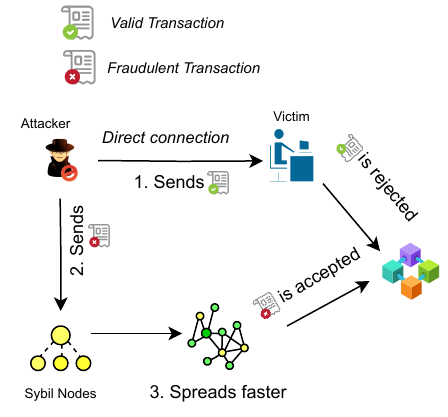}
        \captionof{figure}{Overview of Race Attack}
        \label{fig:race_attack}
    \end{minipage}
    \hfill
    \begin{minipage}[b]{0.49\textwidth}
        \centering
        \includegraphics[width=\textwidth,keepaspectratio]{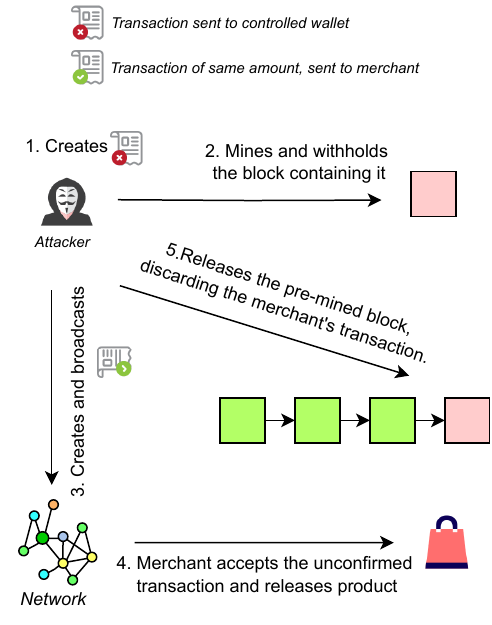}
        \captionof{figure}{Overview of Finney Attack}
        \label{fig:finney_attack}
    \end{minipage}
\end{figure}

\vspace{2mm}
\noindent \textbf{Attack strategy overview:} In this attack, an attacker generates two distinct transactions that utilize the same fund. One transaction goes to the merchant and the other one goes to a wallet controlled by the attacker. Eventually, the merchant releases the product without confirmation. The attack strategy overview is presented in Figure \ref{fig:race_attack}.

\vspace{2mm}
\noindent \textbf{Conditions \& outcomes:} The faster transmission of the fraudulent transaction is necessary for this attack to succeed. In a peer-to-peer system, the attacker must establish a direct connection with the merchant. By doing this, the actual transaction will be sent to the victim merchant more quickly. Moreover, the attacker must also have control over a large number of Sybil nodes to disseminate the fraudulent transaction more quickly than the original one. Thus, a race condition is created. The transaction added to the blockchain is the one that gets to the miners of the network first. The attacker will be able to keep the money and the product if he wins the race. After a successful attack, the merchant suffers from financial loss, and a blockchain fork may occur.

\vspace{2mm}
\noindent \textbf{Enhancements:} This attack does not provide any leverage for subsequent attacks.

\vspace{2mm}
\noindent \textbf{Plausibility \& prevention measures:} If a 0-confirmation payment is accepted anywhere, this attack is possible. Several ways to prevent this attack have been suggested and implemented. For example, adding listening period and observer nodes in network \cite{karame2012two}, reward-based observers \cite{perez2019double}, forwarding double-spending attempts \cite{karame2015misbehavior}, connecting with a large random sample of nodes and not accepting incoming request \cite{bamert2013have} etc. However, complete protection against double-spending attacks is not available yet \cite{bamert2013have}.

\subsection{Finney Attack}
\label{FinneyAttack}
This attack was first suggested by Hal Finney in a bitcoin forum back in 2011 \cite{halFinney}. This is an instance of a double-spending attack where the attacker purposefully spends the same currency multiple times.

\vspace{2mm}
\noindent \textbf{Motivation \& vulnerability:} The attacker wants to double-spend the cryptocurrency. Similar to the Race attack described in Section \ref{RaceAttack}, the attacker targets those merchants who accept `Fast Payment'.

\vspace{2mm}
\noindent \textbf{Attack strategy overview:} The attack strategy overview is presented in Figure \ref{fig:finney_attack}.

\vspace{2mm}
\noindent \textbf{Conditions \& outcomes:} The attacker has to find a valid block first. Additionally, he must verify that the block includes his counterfeit transaction. Merchant also needs to accept 0-confirmation payments. In a successful attack, the attacker receives both the block generation reward and the merchant's product without incurring any additional expenses. The merchant suffers from financial loss. An additional possibility of this attack is the occurrence of a blockchain fork.

\vspace{2mm}
\noindent \textbf{Enhancements:} This attack does not provide any leverage for subsequent attacks.

\vspace{2mm}
\noindent \textbf{Plausibility \& prevention measures:} The complex and time-sensitive nature of this attack makes its occurrence quite improbable. So far, this attack has not been reported for any blockchain system  \cite{saad2020exploring} and hence, it remains only theoretical. No mitigation technique is exclusively available for this attack. However, some general mitigation techniques have been proposed to prevent double-spending attacks described in Section \ref{RaceAttack}.

\subsection{Long Range Attack: Simple}
\label{long_range_simple}
In a Proof-of-Stake (PoS) system, where blocks are generated without the requirement of solving computationally difficult mathematical problems, the possibility of a long-range attack exists \cite{longRangeButerin}. A malicious block validator initiates a process of forking the chain by reverting to the genesis block. He creates a separate branch that has a partially or an entirely different history from the main branch. As the forged branch stretches further than the main branch, previous transactions are eliminated from the chain. Ultimately, the alternate branch is released and the attacker achieves his goal.

\vspace{2mm}
\noindent \textbf{Motivation \& vulnerability:} The main motivation of this attack is to modify the timestamp and manipulate the block history. The Proof of Work (PoW) based systems necessitate a substantial amount of computational power to modify the history of a block. On the contrary, the PoS-based systems utilize a technique known as `Nothing At Stake' \cite{longRangeButerin}. The validator encounters no risk when making consensus decisions. Theoretically, a dishonest validator can construct an alternative branch from the main chain of a Proof of Stake (PoS) blockchain at any desired point, without suffering any tangible costs.

\begin{figure}[!htbp]
    \begin{minipage}[b]{0.48\textwidth}
        \centering
        \includegraphics[width=\textwidth,keepaspectratio]{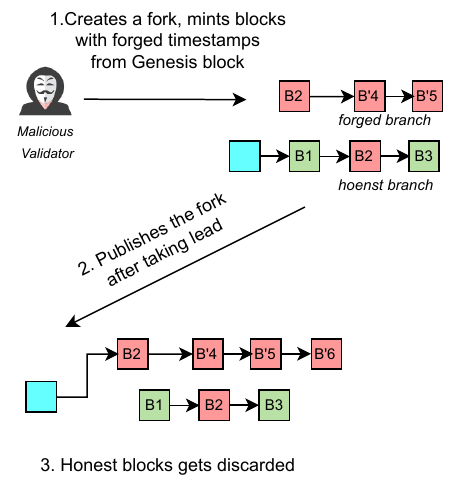}
        \captionof{figure}{Overview of Long Range (Simple) Attack}
        \label{fig:long_range_simple}
    \end{minipage}
    \hfill
    \begin{minipage}[b]{0.48\textwidth}
        \centering
        \includegraphics[width=\textwidth,keepaspectratio]{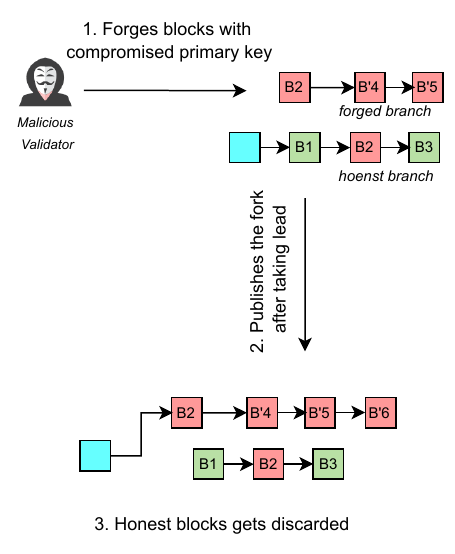}
        \captionof{figure}{Overview of Long Range Attack (Posterior Corruption).}
        \label{fig:long_range_posterior_corruption}
    \end{minipage}
    
\end{figure}

\vspace{2mm}
\noindent \textbf{Attack strategy overview:} The attack strategy overview is presented in Figure \ref{fig:long_range_simple}.

\vspace{2mm}
\noindent \textbf{Conditions \& outcomes:} For this attack to be successful, it is imperative that the block timestamp is disregarded and the longest chain is always selected as the main branch \cite{deirmentzoglou2019survey}. The attacker must generate blocks in advance to gain a competitive advantage. If this attack is successful, the history of blocks becomes corrupted and double-spending is possible. Furthermore, the attack results in the manifestation of `Weak Subjectivity' \cite{deirmentzoglou2019survey}. A Weak Subjectivity refers to the challenge that is faced by newly added nodes and those that are reconnected to the network after a prolonged period of disconnection. Due to their lack of synchronization, these nodes do not possess accurate information regarding the development of the blockchain, which hinders its ability to determine the primary chain.

\vspace{2mm}
\noindent \textbf{Enhancements:} This attack does not provide any leverage for subsequent attacks.

\vspace{2mm}
\noindent \textbf{Plausibility \& prevention measures:} This attack is purely theoretical. No occurrence has been observed till date. This attack can be prevented by considering timestamps while selecting the chain \cite{gavzi2018stake}.

\subsection{Long Range Attack: Posterior Corruption}
\label{long_range_posterior_corruption}
If timestamp forging is not feasible in a PoS system and an attacker possesses the private key of different validators, then this attack can be carried out. Within the alternate branch, the attacker alternates block histories using both his private key and the compromised keys. Following a significant advantage, the attacker deploys the alternative branch of the blockchain, which is subsequently acknowledged and adopted.

\vspace{2mm}
\noindent \textbf{Motivation \& vulnerability:}
The attacker wants to rewrite the block history. He does this with compromised private keys. To maintain an equitable PoS consensus based system, validators must engage in rotation. Additionally, there should be a means to voluntarily or forcibly remove a validator from the system. In a real-world system, a validator may be unreliable due to potential changes in incentives or compromises to the system. Therefore, a validator has the option to retire once they have generated a specific number of blocks (denoted as \textit{n}). He can withdraw his investment from the system by liquidating it. And will no longer be included in it. However, the blocks he created remain within the system. By utilizing his private key, it is possible to fabricate the preceding \textit{n} blocks \cite{poelstraCostless}. The attacker can perhaps engage in bribery with the retired validator or somehow hack his private key. Then, whenever the attacker is elected as a block validator, he possesses the ability to generate further counterfeit blocks by utilizing both their own and the hacked private keys at a faster pace.

\vspace{2mm}
\noindent \textbf{Attack strategy overview:} The attack strategy overview is presented in Figure \ref{fig:long_range_posterior_corruption}. In this case, the attacker has access to the private key of the B1 block validator.

\vspace{2mm}
\noindent \textbf{Conditions \& outcomes:} To make this strategy successful, the attacker needs to control one or more private keys. The outcome of this attack is similar to the simple variant of the same attack described in Section \ref{long_range_simple}.

\vspace{2mm}
\noindent \textbf{Enhancements:} This attack does not provide any leverage for subsequent attacks.

\vspace{2mm}
\noindent \textbf{Plausibility \& prevention measures:}
This attack is not recorded in any blockchain system so far. A few mitigation techniques have been suggested like Frequent checkpoints \cite{gavzi2018stake}, Key Evolving Cryptography \cite{david2018ouroboros}, and Trusted Execution Environment \cite{li2017securing}.

\subsection{Long Range Attack: Stake Bleeding}
\label{long_range_stake_bleeding}
This strategy was proposed by Ga{\v{z}}i \textit{et al.} \cite{gavzi2018stake}. This attack can be launched against PoS consensus-based blockchain systems.

\vspace{2mm}
\noindent \textbf{Motivation \& vulnerability:} The attacker wants to rewrite the history of transactions. In a blockchain system with no frequent checkpoints, absence of context-based transactions, and longest chain rule with transaction fees to be used as a reward mechanism, this attack can be launched. Block validators with any proportion of shares can launch this attack.

\vspace{2mm}
\noindent \textbf{Attack strategy overview:} The attack strategy overview is presented in Figure \ref{fig:long_range_stake_bleeding}.



\vspace{2mm}
\noindent \textbf{Conditions \& outcomes:} This attack has a significant relation with relative stakes. The more stakes the attacker possesses, the less time it requires to complete the attack. So, attackers may form a coalition to launch such an attack \cite{gavzi2018stake}. Additionally, a Stake Bleeding attack could be launched after the prior blockchain has been operational for several years. A successful attack corrupts the block history which can further lead to double-spending.

\vspace{2mm}
\noindent \textbf{Enhancements:} This attack does not provide any leverage for subsequent attacks.

\vspace{2mm}
\noindent \textbf{Plausibility \& prevention measures:} A stake-bleeding attack is difficult to carry out as it would require several years of blockchain history. No such attack has been recorded so far. To effectively launch this attack, an attacker with a 30\% stake would need approximately six years of blockchain history \cite{gavzi2018stake}. However, an Eclipse attack-based stake bleeding attack is proposed by Zhang \textit{et al.} which significantly reduces the attack completion time \cite{zhang2019eclipse}. To prevent this attack, Ga{\v{z}}i \textit{et al.} proposed two methods: Context Sensitive Transaction and Density Detect Mechanism \cite{gavzi2018stake}.

\subsection{P+$\epsilon$ Attack}
\label{P+e_attack}
This attack was first introduced by Vitalik Buterin in \cite{p+epsilon}. Many cryptocurrency systems like SchellingCoin \cite{schelling} operate on the assumption that every participant will act honestly because they believe everyone else will do so. The attacker takes advantage of this assumption and manipulates the user to achieve any desired state.

\vspace{2mm}
\noindent \textbf{Motivation \& vulnerability:} The attacker's motivation is to take over the mechanism at zero cost. He does this by making a promise to reward people who voted in a certain way after the game is over. This reward is only paid if the majority voted differently from what they expected. Because the attacker is only paying out the reward if they win, it is always in the best interest of the voters to vote in the way that the attacker wants, regardless of what they believe the majority will do. The vulnerability of the system is that it relies on the assumption that people will act honestly in a simultaneous consensus game. The attack shows that this assumption is not always valid.

\vspace{2mm}
\noindent \textbf{Attack strategy overview:} Let us assume a system where each participant gets to vote whether to adopt a new consensus mechanism.  A user gets a reward of P if he votes with the majority; 0 otherwise. The reward matrix is given in Table \ref{reward}. In this case, each user has the incentive to vote honestly with the majority.
\begin{table}
\centering
\caption{Reward Matrix for P+$\epsilon$ Attack}
\begin{tabular}{|l|c|c|}
\hline
\textbf{} & \textbf{User Votes 0} & \textbf{User Votes 1}\\
\hline
\textbf{Other Nodes Vote 0} & P & 0 \\
\hline
\textbf{Other Nodes Vote 1} & 0 & P\\
\hline
\end{tabular}
\label{reward}
\end{table}
Now, let us again assume that an attacker wants to change the consensus by manipulating other users. He announces to pay a little extra reward of \(\epsilon\) to those who vote 1 in addition to P after the voting period, if the majority votes against adopting the consensus, that is, the majority votes 0. And if the majority goes with adopting a new consensus, that is, votes 1, nobody gets any extra reward. The reward matrix now will be similar to Table \ref{reward2}. 

In this modified case, the users get manipulated and vote 1. But according to the attacker's contract, if the majority votes 1, nobody gets an extra reward. Ultimately, the attacker loses nothing and achieves the desired state of the system. 

\begin{table}
\centering
\caption{Modified Reward Matrix for P+$\epsilon$ Attack}
\begin{tabular}{|l|c|c|}
\hline
\textbf{} & \textbf{User Votes 0} & \textbf{User Votes 1}\\
\hline
\textbf{Other Nodes Vote 0} & P & P+\(\epsilon\) \\
\hline
\textbf{Other Nodes Vote 1} & 0 & P\\
\hline
\end{tabular}
\label{reward2}
\end{table}
\vspace{2mm}
\noindent \textbf{Conditions \& outcomes:} The attacker needs to broadcast the bribe. It may be done by a smart contract or by giving bribes. A demo Ethereum smart contract is available \cite{peps2018}. By launching this attack, the attacker achieves the desired state of a blockchain system without even paying anything.

\vspace{2mm}
\noindent \textbf{Enhancements:} This attack does not lead to any subsequent attacks.

\vspace{2mm}
\noindent \textbf{Plausibility \& prevention measures:} This is a theoretical attack. Vitalik proposed two strategies \cite{p+epsilon} to mitigate such attacks. The first one is to require users to put down a deposit. Another one is to use counter-coordination strategies.

\subsection{Feather Forking Attack }
\label{FeatherForkingAttack}
The feather forking attack is a more affordable version of the Punitive Forking Attack. It does not require a majority hashpower, similar to Punitive Forking. The first mention of Feather Fork and the strategy of the attack was made on the BitcoinTalk forum \cite{bitcoinTalk_Feather_Forking0}.

\vspace{2mm}
\noindent \textbf{Motivation \& vulnerability:} Feather forking differs from punitive forking in that the majority hashpower is optional, however, the final goal remains the same. It can be used to force victims to pay huge amounts to validate blocks or have their transactions banned but the attacker must have mining pool authority \cite{howtoDestroyPunitive_Forking2,bitcoinTalk_Feather_Forking0}.

\begin{figure*}[!h]  
    \centering
    \includegraphics[width=1.0\textwidth]{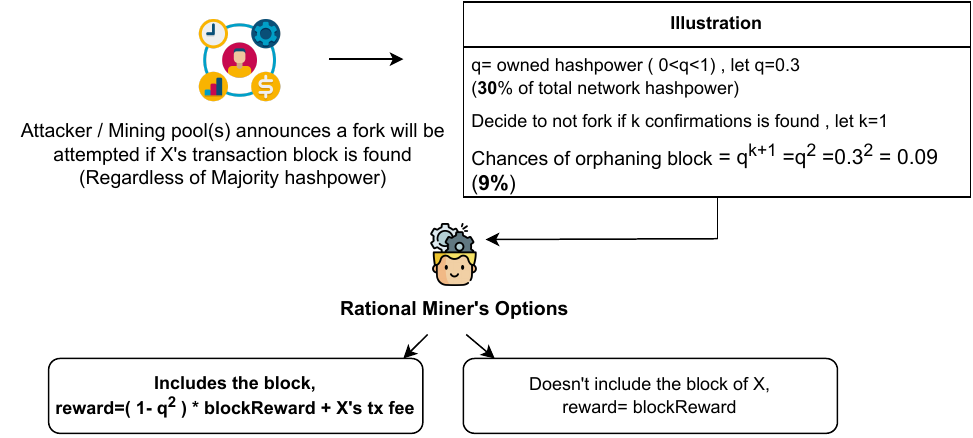}  
    \caption{Overview of Feather Forking Attack}
\label{fig:feather_forking_attack}
\end{figure*}

\vspace{2mm}
\noindent \textbf{Attack strategy overview:} The attack strategy overview is presented in Figure \ref{fig:feather_forking_attack} \cite{magnani2018Feather_Forking1,howtoDestroyPunitive_Forking2}.

\vspace{2mm}
\noindent \textbf{Conditions \& outcomes:} With owning some significant hashpower (<=49\%) , and wagering on new block discovery time, the attacker can enforce significant transaction fee or if attacker discovers some blocks chronologically ( value of K from Figure \ref{fig:feather_forking_attack} ) in forked chain, then the victim's transaction is blacklisted \cite{Hypothetical_at_Feather_Forking2}.

\vspace{2mm}
\noindent \textbf{Enhancements:} This attack does not provide any leverage for subsequent attacks.

\vspace{2mm}
\noindent \textbf{Plausibility \& prevention measures:} All miners in pools utilize reference client programs, therefore this attack would require the majority of the network to use rationally motivated clients, which is unlikely. With the ``RationalMiner" client program, mitigation is still unproposed \cite{bitcoinTalk_Feather_Forking0,Iosr_jeeePunitive_Forking3,conti2018surveyPunitive_Forking1}.

\subsection{Bribery Attack }
\label{BriberyAttack}
A bribery attack is a way to double-spend without owning the majority of hash power and renting the amount of hash power needed from rational miners. The Bribery Attack was first introduced by Joseph Bonneau \cite{bonneau2016buyBribe1} and first implementation was in "Smart Contracts for Bribing Miners" by Mccorry \textit{et al.} \cite{mccorry2018smartBribe2}.

\vspace{2mm}
\noindent \textbf{Motivation \& vulnerability:} The motive is to accquire short lived mining power by leveraging rational miners ( through bribing ) with an aim to gain profit for both attacking individuals and bribed miners. The main chain may face a little less hashpower for a short time, however, it is dependent on the choices of rational miners. And in short bursts, it is hard to detect or mitigate. To incentivize, the attacker must keep rational miners engaged with bribed payments \cite{conti2018surveyPunitive_Forking1,bonneau2016buyBribe1}.

\begin{figure}[!htbp]
    \begin{minipage}[b]{0.48\textwidth}
        \centering
        \includegraphics[width=\textwidth,keepaspectratio]{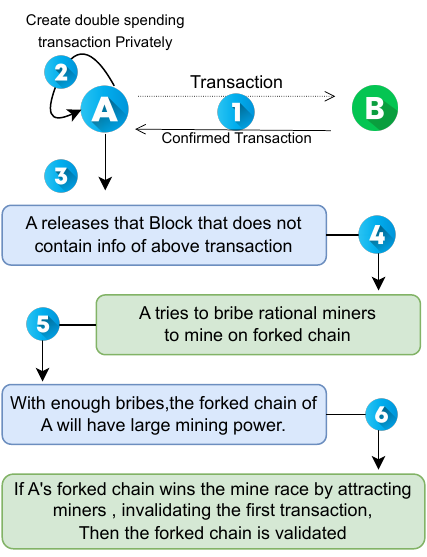}
        \captionof{figure}{Overview of Bribery Attack}
        \label{fig:bribery_attack}
    \end{minipage}
    \hfill
    \begin{minipage}[b]{0.50\textwidth}
        \centering
        \includegraphics[width=\textwidth,keepaspectratio]{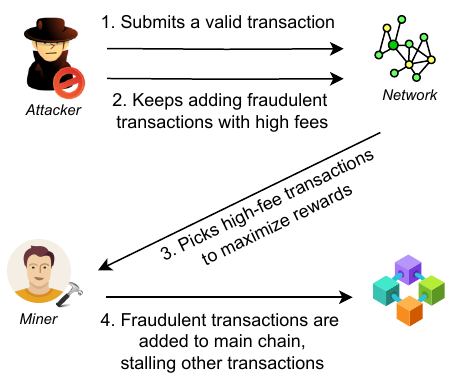}
        \captionof{figure}{Overview of Block Stuffing Attack}
        \label{fig:block_stuffing}
    \end{minipage}
    
\end{figure}

\vspace{2mm}
\noindent \textbf{Attack strategy overview:} The attack strategy overview is presented in Figure \ref{fig:bribery_attack}.

\noindent \textbf{Conditions \& outcomes:} The stakes of attacker are to convince miners for cooperation and join his double-spending cause and proposed methodologies to bribe miners are \cite{bonneau2016buyBribe1} : 
\begin{enumerate}
 \item Out-of-band payment/ Enforced third party mining arrangements
 \item Running a negative-fee mining pool with cash burns to attract miners ( if miner's don't stick, there's no established financial trust )
 \item In-band payment via forking, create a soft fork to test if miners adopt this with funds available/ announced on the forked chain . A fork failure does not waste money, hence this is safer.
\end{enumerate}
For a strong attack, it strongly encouraged to own around 15\% hashrate. This allows participation in the block race after publishing the double-spending block \cite{ebrahimpour2021analysisMainBribe0}, resulting in nearly 90\% relative gains, as quantified by Sun \textit{et al} \cite{sun2020modelBribe3}.

\vspace{2mm}
\noindent \textbf{Enhancements:} This attack does not provide any leverage for subsequent attacks.

\vspace{2mm}
\noindent \textbf{Plausibility \& prevention measures:} The attack is plausible with good strategies, but there is no precedent. After detection, victims / miners can counter-bribe, although it is not certain to work \cite{bonneau2016buyBribe1}.
Attempting to bribe individual miners requires a large budget and tolerance. Other solutions include fund transfer limits and block limited transfers \cite{ebrahimpour2021analysisMainBribe0}. Another mitigation approach is to increase the amount of confirmation blocks \cite{sun2020modelBribe3}.

\subsection{Consensus delay attack}
\label{ConsensusDelayAttack}
This attack, which causes temporary block and transaction delays, has been mentioned in several studies \cite{10.1145/2810103.2813655ConsensusDelay01}. The latency of system consensus is the time it takes to reach consensus. This attack aims to postpone consensus and reveal the PoW consensus constraints on node synchronization and time limits \cite{194906ConsensusDelay00}. This attack may occur before a full-scale attack.

\vspace{2mm}
\noindent \textbf{Motivation \& vulnerability:} The consensus attack exploits the mechanisms of block verifications and authenticity checks in order to exploit the time required for blockchains to achieve consensus. This, in turn, causes delays in targeted transactions by impeding the propagation of blocks. In Bitcoin, the PoW consensus system implemented by Nakamoto includes a timeout of 20 minutes for receiving updated information on newly found blocks (inv messages \cite{bitcoin_struct_consensusdelay02}). This timeout is being exploited in this context. In different consensus systems, there are many methods available to postpone the dissemination of information in order to achieve the same objective \cite{10.1145/2810103.2813655ConsensusDelay01,castro1999practical}. 

\begin{figure*}[!h]  
    \centering
    \includegraphics[width=1.0\textwidth]{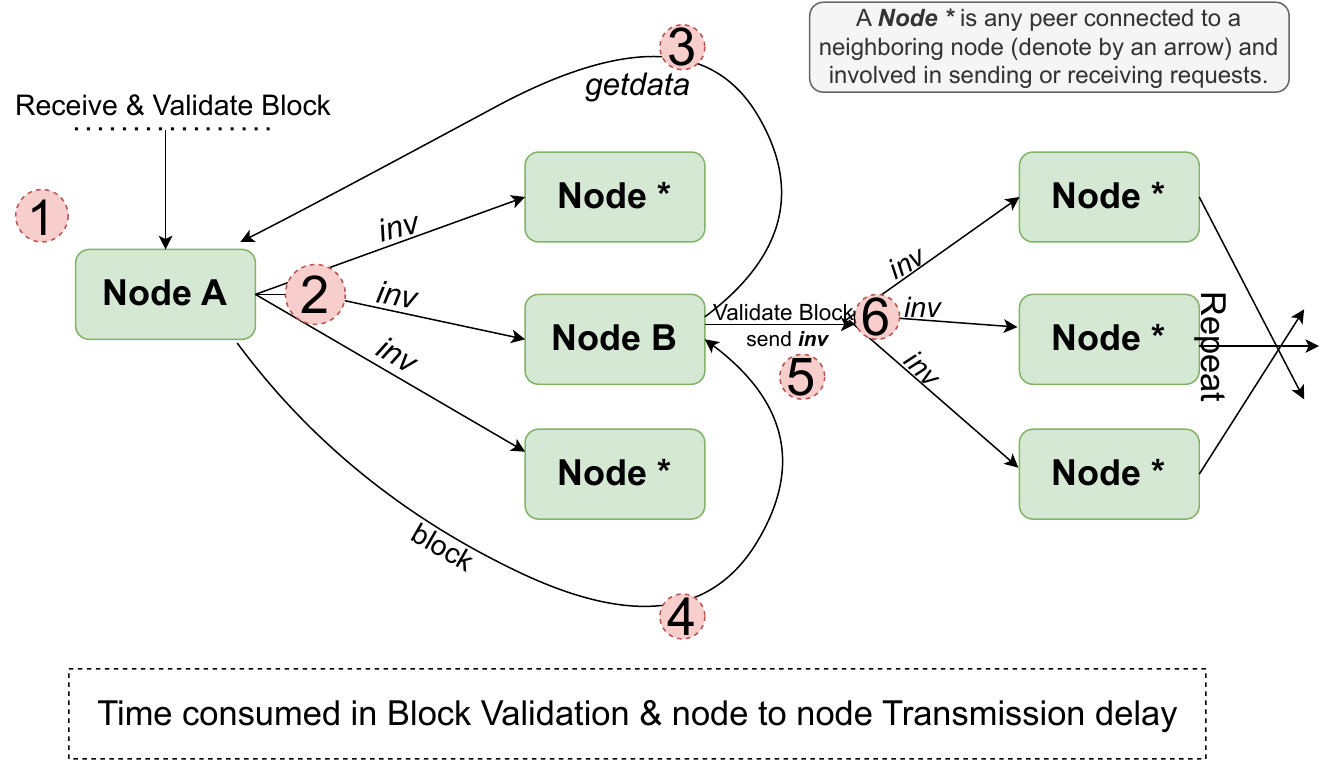}  
    \caption{Overview of Consensus Delay Attack}
\label{fig:consensus_delay_attack}
\end{figure*}

\vspace{2mm}
\noindent \textbf{Attack strategy overview:} The block management overview (which is exploited in Bitcoin) is presented in Figure \ref{fig:consensus_delay_attack} \cite{10.1145/2810103.2813655ConsensusDelay01}.

\begin{figure*}[!h] 
   \centering
        \includegraphics[width=1\textwidth,keepaspectratio]{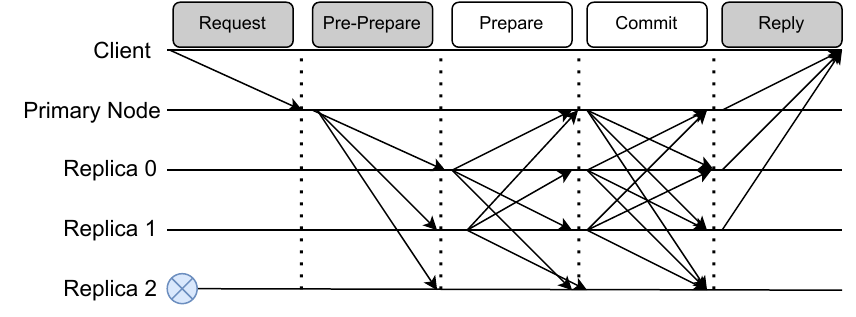}
        \captionof{figure}{PBFT Consensus Algorithm Process}
        \label{fig:consensus_delay_of_pbft_attack}
\end{figure*}

While Bitcoin exploits the previous block management overview, some distributed systems and platforms, such as Zilliqa \cite{Zilliqa}, Hyperledger Fabric \cite{hyperledger}, and Tendermint \cite{tendermint}, utilize the Practical Byzantine Fault Tolerance (PBFT) consensus algorithm \cite{castro1999practical}. PBFT is employed in private and permissioned blockchains to achieve consensus, even in the presence of a small number of faulty nodes.

In PBFT, the primary node with other backup nodes executes a 4 steps algorithm to confirm a transaction.

 During Prepare and Commit Phase (Figure \ref{fig:consensus_delay_of_pbft_attack}) :
 \begin{itemize}
   \item Control a few replicas and delay introduction is possible \cite{zhong2023byzantine}.
   \item Send bogus signatures to other replicas.
 \end{itemize}

\vspace{2mm}
\noindent \textbf{Conditions \& outcomes:} First, the attacker must be a full-node, not SPV or other clients. A full node maintains the entire blockchain, independently verifying all transactions, while an SPV (Simplified Payment Verification) node relies on partial data and external sources for transaction validation \cite{10.5555/2695500SybilAttack04}.  Attackers often control numerous sybil nodes since delaying information to other nodes is difficult. Timing is critical for targeting node transaction delays. If the strategy works, genuine blocks will be wasted and honest miners would lose crucial time. With higher hashpower (33\% advocated by Eyal \textit{et al.} \cite{eyal2014majority}), selfish mining attacks could occur simultaneously.

\vspace{2mm}
\noindent \textbf{Enhancements:} Successful attacks creates leverage for Sybil Attack, 51\% Attack, DoS attack and Double-Spending \cite{10.1145/2810103.2813655ConsensusDelay01}.

\vspace{2mm}
\noindent \textbf{Plausibility \& prevention measures:} Strategies make it viable, and suggested mitigation methods are additional relay network data, Change inv \cite{bitcoin_struct_consensusdelay02} messages and transaction advertisers amd Non-responders penalty \cite{10.1145/2810103.2813655ConsensusDelay01}.

\section{Application Layer Attacks}
\label{sec:appLayerAttack}
The application layer describes how we engage with the blockchain by reading its data and enables operations such as cryptocurrency creation, cryptocurrency distribution, smart contract deployment, and even incentive systems. Application layer attacks are especially directed at these functionalities. Unlike attacks that directly target the main blockchain, these attacks leverage flaws in the blockchain applications, potentially causing catastrophic repercussions.

\subsection{Replay Attack }
\label{ReplayAttack}

Replay attacks are common on blockchains. This exploit tries to spend transaction data from one chain on another legal split chain.
Hard forks, which change or improve blockchain systems, facilitate the ability to replay attacks \cite{coinSutraReplay1}.

\vspace{2mm}
\noindent \textbf{Motivation \& vulnerability:} After a hard fork on blockchain, the attacker targets to replay transactions from captured/sniffed/collected data. This attack in financial organizations can duplicate transactions and steal funds from unsuspecting accounts \cite{Jacalynn_2022}.

\begin{figure}[!h]
    \begin{minipage}[b]{0.46\textwidth}
        \centering
        \includegraphics[width=\textwidth,keepaspectratio]{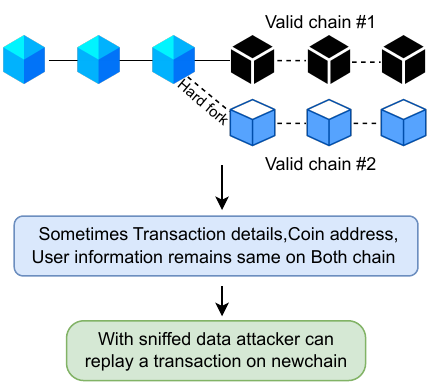}
        \captionof{figure}{Overview of Replay Attack}
        \label{fig:replay_attack}
    \end{minipage}
    \hfill
    \begin{minipage}[b]{0.50\textwidth}
        \centering
        \includegraphics[width=\textwidth,keepaspectratio]{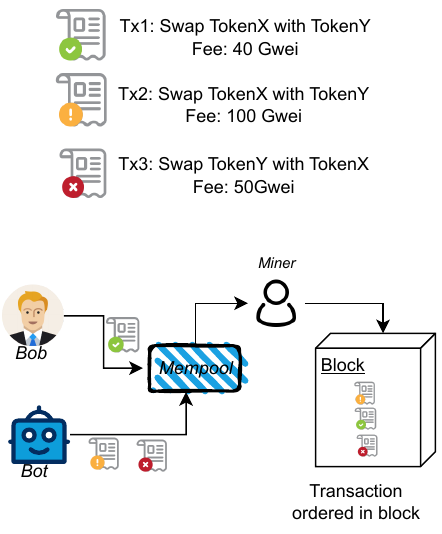}
        \captionof{figure}{Overview of Front Running Attack}
        \label{fig:front_run_sandwich}
    \end{minipage}
\end{figure}

\vspace{2mm}
\noindent \textbf{Attack strategy overview:} The attack strategy overview is presented in Figure \ref{fig:replay_attack} \cite{coinSutraReplay1}.

\vspace{2mm}
\noindent \textbf{Conditions \& outcomes:} The attack depends on the replay timing. The attacker must attempt the transaction on the new/updated chain with correct data after the hard fork and before replay protection on the upgraded chain. A successful execution could earn funds in this way \cite{cryptocurrencyfactReplay2}.

\vspace{2mm}
\noindent \textbf{Enhancements:} This attack does not provide any leverage for subsequent attacks.

\vspace{2mm}
\noindent \textbf{Plausibility \& prevention measures:} System improvements have caused hard forks, making the attack possible with timing. However, replay protection, time stamping, and unique transaction attributes have been shown to minimize this \cite{coinSutraReplay1,cryptocurrencyfactReplay2}.

\subsection{Short-Address Attack }
\label{Short-AddressAttack}

This is not a full-scale attack, but rather a representation of a minor vulnerability found in a third-party application that communicates with Solidity contracts \cite{mediumShortAddress0}. \cite{mediumShortAddress0}.

\vspace{2mm}
\noindent \textbf{Motivation \& vulnerability:} The bug's that existed in the Contract ABI Specification \cite{Contract_ABI} allowed to send shorter parameters than expected, which allowed stealing tokens from exchanges. The main factor is address generations, which leads to cryptocurrency inflation \cite{ShortAddress1}.

\vspace{2mm}
\noindent \textbf{Enhancements:} This attack does not provide any leverage for subsequent attacks.

\vspace{2mm}
\noindent \textbf{Plausibility \& prevention measures:} Once detected, the flaw was fixed, making the attack impossible in the present time \cite{mediumShortAddress0}.

\section{Meta Application Layer Attacks}
\label{sec:metLayerAttack}
The Meta-application layer attacks in a blockchain system target the overlay that leverages the semantic interpretation of blockchain for a range of application domains. These attacks target the semantics of the blockchain system and the functionality beyond specific applications. The meta-application layer attacks exploit flaws inherent in decentralized applications. Decentralized applications or DApps are programs that run on a blockchain or p2p network. These apps allow direct user interactions without any intermediaries. They use smart contracts to automate transactions and processes, eliminating the need for intermediaries. OpenSea \cite{open2017sea}, Ethereum Name Service \cite{ens2017}, Cryptokitties \cite{cryptokitties} etc. are some of the examples of such apps. Decentralized Exchanges or DExs are examples of meta-applications. These are cryptocurrency trading platforms based on blockchain network. DExs, unlike traditional exchanges, are not centralized, strengthening their resistance to manipulation and censorship. Users can trade directly with one another through smart contracts, which ensures transparency and security. Uniswap \cite{uniswap2020}, Sushiswap \cite{sushi2020swap}, Bancor \cite{bancor2017} etc. are some famous DExs. The attacks on this layer can breach the integrity, security, and interoperability of blockchain-based services across various application domains.

\subsubsection{Front Running Attack}
\label{front_running}

Front-running is a term associated with the share market. It refers to a scenario where a broker has confidential information about a large upcoming transaction that will significantly influence the price of the associated shares. Consequently, the attacker buys the share before it becomes publicly available. This is considered ill-practice \cite{frontrun}. Eskandari \textit{et al.} organized a front-running attack associated with the blockchain-based system in four different categories: decentralized exchanges, crypto-collectible games, naming services and gambling apps \cite{eskandari2019sok}.

\vspace{2mm}
\noindent \textbf{Motivation \& vulnerability:} Front-running attacks in blockchain are motivated by financial gain. With knowledge of impending transactions, attackers influence the system to secure profit from price manipulation, arbitrage and fee extraction. DApps are vulnerable to such attacks. Most blockchains have public mempools with visible pending transactions before confirmation. This transparency lets attackers learn about upcoming events. Also, faulty smart contracts can cause such an attack \cite{solidityfrontrun}. Maximum Extractable Value (MEV) bots play a vital role in front-running attacks. These bots are operated by validators or independent actors. They strategically reshuffle, include, or remove mempool transactions to maximize value \cite{frontruncomprehensive}.

\vspace{2mm}
\noindent \textbf{Attack strategy overview:} Many cryptocurrency exchanges have bots. They constantly search the mempool for transactions and purchases as necessary. Let us assume a case where an honest user, Bob, wants to swap a significant amount of TokenX with TokenY. The demand-supply theory says this transaction will raise TokenY's price. A malicious bot observes this and broadcasts a similar transaction, but a slightly smaller amount of token. In this transaction, the bot attaches a slightly more gas fee. Another transaction by the bot swaps TokenY for TokenX, reducing its price. This time, the bot attaches a slightly lower gas fee. Miners will arrange these transactions in the ascending order of gas fee. So, Bob will ultimately gain less than expected.

An overview of the attack strategy is presented in Figure \ref{fig:front_run_sandwich}.

\vspace{2mm}
\noindent \textbf{Conditions \& outcomes:} The attacker needs a total knowledge of market and transaction history. This type of attack disrupts the market or provides an unfair competition advantage. Also, such attacks may lead to a double-spending scenario \cite{eskandari2019sok}.

\vspace{2mm}
\noindent \textbf{Enhancements:} This attack does not provide any leverage for subsequent attacks.

\vspace{2mm}
\noindent \textbf{Plausibility \& prevention measures:} 
This attack is a very common issue in Ethereum DApps or in any DEx. DODO DEx suffered \$3.8 million loss \cite{dodo2020} and bZx DEx suffered a massive loss of \$350,000 in such an attack \cite{bzx}. Several mitigation techniques such as transaction sequencing, confidentiality, and improved design practices \cite{eskandari2019sok} have been proposed to prevent such attacks.

\subsection{Block Stuffing Attack}
\label{block_stuffing_attack}
This is a Denial-of-Service (DoS) attack where the attacker repeatedly sends fake transactions to the network, thus slowing future transactions from being added to the chain. We study this attack in light of a real-life event. FOMO3D \cite{FOMO3D} is a famous gambling game that can be played via an Ethereum smart contract. In this game, players buy keys from a smart contract and deposit money in a pot. Each round begins with a 24-hour time counter. A key purchasing event adds 30 seconds to the counter. When the counter strikes 0, the last person to buy a key wins most of the pot and the rest is shared among the players.

\vspace{2mm}
\noindent \textbf{Motivation \& vulnerability:} The main motivation of such an attack is to stall the network. In case of FOMO3D, the attacker took advantage of the block gas limit and launched this attack. Each block has a certain amount of gas limit. Miners include high-fee transactions to increase profit. Thus, they selected a set of transactions that maximizes profit per block. An attacker's main purpose is to manipulate the selection process by creating a set of transactions with precise calculations that have the best likelihood of being mined to deplete blocks' gas limits and block other transactions from being added to the chain.

\vspace{2mm}
\noindent \textbf{Attack strategy overview:} The attack strategy overview is presented in Figure \ref{fig:block_stuffing}.

\vspace{2mm}
\noindent \textbf{Conditions \& outcomes:} Each transaction incurs a gas price that the attacker must pay. If the attack fails, he will suffer financial losses. The outcome is context-dependent. In case of FOMO3D, the attacker won 10,469 ETH in the first round and 3264.668 ETH in the second round which is equivalent to 24,978,405.86\$ and 7,768,440.74\$ respectively as of February 2024. 

\vspace{2mm}
\noindent \textbf{Enhancements:} This attack creates no leverage for another attack.

\vspace{2mm}
\noindent \textbf{Plausibility \& prevention measures:} The Ethereum network experienced multiple instances of this attack in 2018 \cite{onur, secbit2018}. For developers of smart contracts, several preventive steps have been suggested to avoid smart contract issues and transaction-blocking events at the network level for games and applications of a similar nature like FOMO3D \cite{secbit2018}.

\section{Discussion}
\label{discussion} 
In this section, we provide a summary of our findings from the analysis of a number of attacks (Section \ref{Findings}) and a comparative analysis of our analysis with similar existing research works (Section \ref{subsec:compA}).

\subsection{Summary of Layer-based Attack Analysis}
\label{Findings}
\begin{figure}[h!]
        \vspace*{0.9cm}
        \hspace*{-0.7cm}  
        \centering
	    \includegraphics[width=1\textwidth]{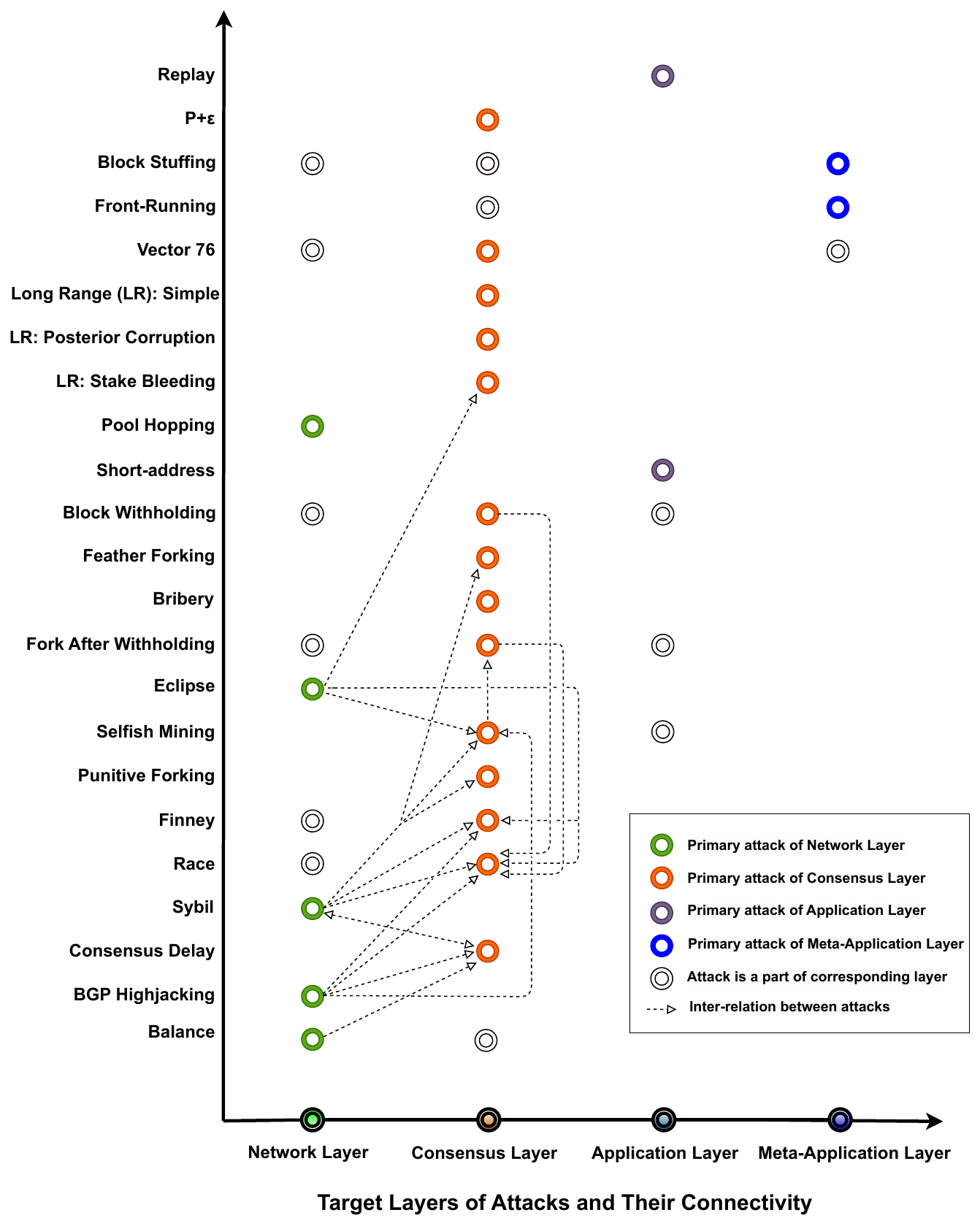}
        \caption{Distribution of studied attacks in different layers}
        \label{fig:layer_based_attack_distribution}     
    \end{figure}

In this article, we have explored several attacks related to blockchain-based systems. We also organize them using a layer-based modeling technique. The inter-relation of these attacks is presented in Figure \ref{fig:layer_based_attack_distribution}

In Figure \ref{fig:layer_based_attack_distribution}, we have placed our studied attacks on the vertical axis (denoted circles) with the corresponding layers on the horizontal axis. If an attack impacts multiple layers, we have pointed them out in the figure as explained by the legend in Figure \ref{fig:layer_based_attack_distribution}. For an attack spanning multiple layers, the primary/targeted layer is denoted with a colorful circle. Different colorful circles have different semantics. These semantics have been added as legends in Figure Figure \ref{fig:layer_based_attack_distribution}. Each arrow connecting the circles suggests a potential scenario of attack enhancement. The starting of the arrow denotes the attack from which the leverage is obtained, while the point of the arrow shows where this leverage can be applied to initiate a new attack.

From our analysis, it is evident that:
\begin{itemize}
\item The network layer is the initial vulnerable layer on the blockchain attack surface.

\item Most attacks aim to exploit the consensus layer to cause significant damage.

\item Some of the attacks (i.e.- BGP Hijacking, Sybil, Eclipse) are very lucrative for attackers as they provide unparalleled advantage for further attacks by enhancement. And all of them targets the network layer primarily.

\item Moreover, few attacks like Consensus Delay, Race \& Finney are most vulnerable by enhancement from other attacks.
\end{itemize}

We also present a summary of the affected layers in Table \ref{tab:layer-table}. Here, we describe the properties of each layer that is exploited in order to make an attack successful for that particular layer. We have the following findings.
\begin{itemize}
\item The consensus layer has the most exploitable properties as almost all of our studied attacks take advantage of protocols or rules of this layer.
\item The network layer is also very crucial because many attacks have to subvert or circumvent this layer to affect the next layer.
\item The application and meta-application layers are crucial for blockchain security, as high-level development involving various API gateways and architectures is prone to vulnerabilities. Bugs in these layers can cause significant financial losses and erode trust in the blockchain. Ensuring the security of these layers is imperative, as they are the primary interface for users in decentralized applications (DApps) and similar platforms, where critical issues can arise.
\end{itemize}

Table \ref{tab:my-table} provides a summary of our findings on individual attacks based on the taxonomy outlined in Section \ref{taxonomy}. In this table, symbols such as `$\varnothing$' and `?' are used to indicate a lack of information or insufficient data for that particular property. On the other hand, symbols such as `$\CIRCLE$', `$\Circle$' and `$\RIGHTcircle$' are used to indicate the presence, absence and partially presence with respect to a particular property. For other properties, descriptive texts have been provided for clarification.

Our findings from Table \ref{tab:my-table} are the following.

\begin{itemize}
\item Of all the attacks, the balance attack is the only attack without any mitigation proposal \& can be executed with the lowest mining power.
\item Except for the Balance attack, all other studied attacks have at least one mitigation proposal. For Replay \& Short-address the solution has already implemented.
\item For Eclipse \& Sybil attacks, the attack surface is complex, and prevention methods are continually being developed. Some of the proposed methods are implemented.
\item Most of the attacks require the motivation of a single adversary miner. In some cases, agents like Sybil nodes help to facilitate the attack faster.
\end{itemize}

\clearpage
\vspace*{1.8cm}
\newcolumntype{C}[1]{>{\centering\arraybackslash}p{#1}}
{\small  
\captionsetup{type=table}
\caption{Layer-based attack categorization and exploiting points for each layer}
\label{tab:layer-table}
\begin{longtable}[H]{|C{2.3cm}|C{2.8cm}|C{2.8cm}|C{2.5cm}|C{1.7cm}|}

\hline
\rowcolor[HTML]{F0EBEB} 
{\color[HTML]{000000} \textbf{Attack Name}} &
  {\color[HTML]{000000} \textbf{Network Layer}} &
  {\color[HTML]{000000} \textbf{Consensus Layer}} &
  {\color[HTML]{000000} \textbf{Application Layer}} &
  {\color[HTML]{000000} \textbf{Meta-Application Layer}} \endhead \hline
  Balance Attack &
  Leveraging knowledge of miner's network structures, computational power, and mining difficulty to introduce communication delays. &
  Leverages Ethereum's GHOST Protocol / Bitcoin's longest chain rule.
  &
   &
   \\ \hline
\rowcolor[HTML]{EEEEEE} 
BGP Hijacking &
  Manipulating internet routing table &
   &
   &
   \\ \hline
Bribery Attack &
   &
  Manipulating miners through bribing &
   &
   \\ \hline
\rowcolor[HTML]{EEEEEE} 
Consensus Delay &
   &
  Exploits the mechanisms of block verifications and authenticity checks &
   &
   \\ \hline
Eclipse Attack &
  P2P Protocol limitations &
   &
   &
   \\ \hline
\rowcolor[HTML]{EEEEEE} 
Sybil Attack &
  Capture new nodes (assisted by multiple devices, VM, IP) &
   &
   &
   \\ \hline
Feather Forking &
   &
  Exploit by forks (using Majority hashpower) &
   &
   \\ \hline
\rowcolor[HTML]{EEEEEE} 
Punitive Forking &
   &
  Exploits by forks (without Majority hashpower) &
   &
   \\ \hline
Replay Attack &
   &
   &
  Absence of replay protection on the upgraded chain &
   \\ \hline
\rowcolor[HTML]{EEEEEE} 
Short-Address &
   &
   &
  Third Party API Exploit &
   \\ \hline
Selfish Mining &
   &
  PoW mechanism exploit by hiding blocks & Reward mechanism exploit &
   \\ \hline
\rowcolor[HTML]{EEEEEE} 
Block Withholding &
  Mining Pool Infiltration &
  PoW mechanism exploit by hiding blocks or avoiding block submission & Reward mechanism exploit &
   \\ \hline
Fork After Withholding &
  Mining Pool Infiltration &
  PoW mechanism exploit by hiding blocks & Reward mechanism exploit &
   \\ \hline
\rowcolor[HTML]{EEEEEE} 
Block Stuffing &
   Denial of Service &
  PoW mechanism exploit by using higher trx fee &
   &
  FOMO3D Application exploit \\ \hline
Finney Attack & Block propagation manipulation &
  PoW mechanism exploit by replacing 0-confirmation transaction &
   &
   \\ \hline
\rowcolor[HTML]{EEEEEE} 
Race Attack &
  Block propagation manipulation &
  PoW mechanism exploit by replacing 0-confirmation transaction &
   &
   \\ \hline
Vector 76 &
  P2P network joining manipulation &
  PoW mechanism exploit by replacing 1-confirmation transaction & Crypto Exchange application exploit &
   \\ \hline
\rowcolor[HTML]{EEEEEE} 
Long Range: Simple &
   &
  PoS mechanism exploit by rewriting block history &
   &
   \\ \hline
Long Range: Posterior Corruption &
   &
  PoS mechanism exploit by rewriting block history &
   &
   \\ \hline
\rowcolor[HTML]{EEEEEE} 
Long Range: Stake Bleeding &
   &
  PoS mechanism exploit by rewriting block history &
   &
   \\ \hline
P+Epsilon &
   &
  Consensus mechanism exploit through bribing user &
   &
   \\ \hline
\rowcolor[HTML]{EEEEEE} 
Front Running &
   &
 Consensus mechanism through transaction reordering &
   &
  DNS, Gambling, DeX etc. application exploit\\ \hline
  Pool Hopping &
 Timely join \& exit mining pools based on profitability and anticipated behavior & &
   &
   \\ \hline
\end{longtable}
}
\addtocounter{table}{-1}

\newpage 
\newcolumntype{C}[1]{>{\centering\arraybackslash}p{#1}}

{\small 
\captionsetup{
  type=table,
  justification = centering
}
\caption{
Taxonomical properties of studied attacks
}
\label{tab:my-table}

\begin{longtable}{@{}|C{2.0cm}|C{2.1cm}|C{1.6cm}|C{1.5cm}|C{1.0cm}|C{1.0cm}|C{2.2cm}|@{}}
\toprule
\rowcolor[HTML]{F0EBEB} 
{\color[HTML]{000000} \textbf{Attack Name}} &
  {\color[HTML]{000000} \textbf{Initiator}} &
  {\color[HTML]{000000} \textbf{Mining Power Requirement\phantom{a}( Minimum / Advised )}} &
  {\color[HTML]{000000} \textbf{Initial Breach Layer}} &
  {\color[HTML]{000000} \textbf{Miti-gation Techniques}} &
  {\color[HTML]{000000} \textbf{Enhanc-ements}} &
  {\color[HTML]{000000} \textbf{Reference}} \endhead \midrule

Balance Attack &
  Single Miner &
  5\% &
  Network &
  \textbf{?} &
  \CIRCLE & \cite{8023156BALANCE00,app9091788BALANCE01}\\ \midrule
\rowcolor[HTML]{EEEEEE} 
BGP Hijacking &
  Single Miner &
  $\varnothing$ &
  Network &
  \RIGHTcircle &
  \CIRCLE & \cite{apostolaki2017hijackingBGP1,saad2022routechainBGP3,sentana2021blockjackBGP4,mastilak2020enhancingBGP6}\\ \midrule
Bribery Attack &
  Single Miner &
  15\% &
  Consensus &
  \RIGHTcircle &
  \Circle & \cite{bonneau2016buyBribe1,ebrahimpour2021analysisMainBribe0,sun2020modelBribe3}\\ \midrule
\rowcolor[HTML]{EEEEEE} 
Consensus Delay &
  Single (Agents) &
  $\varnothing$ &
  Consensus &
  \RIGHTcircle &
  \CIRCLE & \cite{10.1145/2810103.2813655ConsensusDelay01,bitcoin_struct_consensusdelay02,eyal2014majority}\\ \midrule
Eclipse Attack &
   Single (Agents), Groups &
  $\varnothing$ &
  Network &
  \CIRCLE &
  \CIRCLE & 
   \cite{heilman2015Eclipse01,karl2016ethereumEclipse02,marcus2018lowEclipse04,eclipseEclipse03,zhang2019eclipse,Heilman_2015}\\ \midrule
\rowcolor[HTML]{EEEEEE} 
Sybil Attack &
  Single (Agents) &
  $\varnothing$ &
  Network &
  \CIRCLE &
  \CIRCLE & \cite{swathi2019preventingSybilAttack00,ImpervaSybilAttack,10.1007/3-540-45748-8_24SybilAttack01,10.5555/2695500SybilAttack04,BinanceAcademy_2020}\\ \midrule
Feather Forking &
  Pools &
  \textless{}51\% &
  Consensus &
  \RIGHTcircle &
  \Circle & \cite{bitcoinTalk_Feather_Forking0,Iosr_jeeePunitive_Forking3,conti2018surveyPunitive_Forking1,howtoDestroyPunitive_Forking2,Hypothetical_at_Feather_Forking2}\\ \midrule
\rowcolor[HTML]{EEEEEE} 
Punitive Forking &
  Pools &
  51\% &
  Consensus &
  \RIGHTcircle &
  \Circle & \cite{miningPoolsSummaryBest_2024,conti2018surveyPunitive_Forking1,Punitive_Forking0} \\ \midrule
Replay Attack &
  Single Miner &
  $\varnothing$ &
  Application &
  \CIRCLE &
  \Circle & \cite{coinSutraReplay1,cryptocurrencyfactReplay2}\\ \midrule
\rowcolor[HTML]{EEEEEE} 
Short-Address &
  Single Miner &
  $\varnothing$ &
  Application &
  \CIRCLE &
  \Circle & \cite{mediumShortAddress0,ShortAddress1}\\ \midrule
Selfish Mining &
  Single Miner &
  9\% &
  Consensus &
  \RIGHTcircle &
  \CIRCLE & \cite{solat2016zeroblock,saad2019countering,kwon2017selfish}\\ \midrule
\rowcolor[HTML]{EEEEEE} 
Block Withhelding &
  Single Miner,  Groups, Pools &
  $\varnothing$ &
  Network &
  \RIGHTcircle &
  \CIRCLE & \cite{wizEligius,rosenfeld2011analysis, courtois2014subversive,eyal2015miner}\\ \midrule
Fork After Withhelding &
  Single Miner &
  $\varnothing$ &
  Network &
  \RIGHTcircle &
  \CIRCLE & \cite{rosenfeld2011analysis, eyal2014disincentivize, kwon2017selfish}\\ \midrule
\rowcolor[HTML]{EEEEEE} 
Block Stuffing &
  Single ( User ) &
  $\varnothing$ &
  Consensus &
  \RIGHTcircle &
  \Circle & \cite{onur, secbit2018}\\ \midrule
Finney Attack &
  Single Miner &
  $\varnothing$ &
  Consensus &
  \RIGHTcircle &
  \Circle & \cite{saad2020exploring}\\ \midrule
\rowcolor[HTML]{EEEEEE} 
Race Attack &
  Single ( User ) &
  $\varnothing$ &
  Network &
  \RIGHTcircle &
  \Circle & \cite{gavzi2018stake,bamert2013have,karame2012two}\\ \midrule
Vector 76 &
  Single Miner &
  $\varnothing$ &
  Application &
  \RIGHTcircle &
  \Circle & \cite{vector76}\\ \midrule
\rowcolor[HTML]{EEEEEE} 
Long Range :Simple &
  Single ( Validator ) &
  $\varnothing$ &
  Consensus &
  \RIGHTcircle &
  \Circle & \cite{longRangeButerin,deirmentzoglou2019survey}\\ \midrule
Long Range :Posterior Corruption &
  Single ( Validator ) &
  $\varnothing$ &
  Application &
  \RIGHTcircle &
  \Circle & \cite{gavzi2018stake,david2018ouroboros,li2017securing}\\ \midrule
\rowcolor[HTML]{EEEEEE} 
Long Range :Stake Bleeding &
  Single ( Validator ) &
  $\varnothing$ &
  Consensus &
  \RIGHTcircle &
  \Circle & \cite{zhang2019eclipse,gavzi2018stake}\\ \midrule
P+Epsilon &
  Single ( User ) &
  $\varnothing$ &
  Consensus &
  \RIGHTcircle &
  \Circle & \cite{p+epsilon,peps2018}\\ \midrule
\rowcolor[HTML]{EEEEEE} 
Front Running &
  Single ( User ) &
  $\varnothing$ &
  Meta - Application &
  \RIGHTcircle &
  \Circle & \cite{eskandari2019sok,frontruncomprehensive}\\ \midrule
  Pool Hopping &
  Single Miner &
  $\varnothing$ &
  Network &
  \RIGHTcircle &
  \Circle & \cite{singh2019smartPoolHopping1,rosenfeld2011analysis,PoolHopping3,9186838PoolHopping5}\\ \bottomrule
\end{longtable}%
}
\addtocounter{table}{-1}

\subsection{Comparative Analysis}
\label{subsec:compA}
In Table \ref{tab:comparison_of_works} we compare this work with previous survey of attacks on blockchain systems based on distinct properties such as number of attacks covered, number of layers, inter-relation of attacks between layers, mitigation techniques, attacker's perspective analysis and layer vulnerability analysis. In this table, the symbols `$\CIRCLE$',`$\Circle$' and `$\RIGHTcircle$' have been used to indicate the presence, absence and partially presence of the corresponding property respectively. This table distinguishes our work from previous studies in the following manner:
\begin{itemize}
    \item Our attack categorization consists of four layers, closely aligning with \cite{homoliak2020security}. However, in \cite{homoliak2020security}, Homoliak \textit{et al.} employs a Replication State Machine layer, we instead utilize an Application layer and a Meta-Application layer. As a result, the categorization in this work is more refined compared to the four-layer model of \cite{homoliak2020security} and more concise than the six-layer model presented in \cite{wen2021attacks}. Furthermore, our layer categorization is platform-agnostic, unlike \cite{chen2020survey}, which is specifically tailored for Ethereum. Therefore, in terms of layer vulnerability analysis, we also differ from \cite{homoliak2020security} and \cite{wen2021attacks}. This analysis is missing in \cite{saad2020exploring}, \cite{guggenberger2021structured}, \cite{li2020survey}, \cite{zhang2019security} and \cite{wen2021attacks}.
    
     \item We present a detailed examination of the inter-relationship of attacks between layers in terms of our custom framework presented in Section \ref{taxonomy}. For example, how an attack might lead to subsequent attacks, how attacks can affect the components of several levels at the same time, and so on. This detailed examination is not present in previous works.
    \item We discuss mitigation techniques for each attack we studied which are not present in \cite{guggenberger2021structured} and partially discussed in \cite{li2017securing}, \cite{chen2022survey}, \cite{chen2020survey}, \cite{moubarak2018blockchain}. While \cite{anita2019blockchain} points out some mitigation techniques, our work covers a broader range of attacks. Our discussion closely aligns with the approaches found in \cite{homoliak2020security}, \cite{saad2020exploring}, and \cite{wen2021attacks}.

    \item We provide an in-depth analysis from the attacker's perspective, detailing which vulnerabilities in specific layers are exploited, the motivations behind these attacks, and the associated stakes of launching an attack. This analysis is missing in most of the aforementioned previous works and only briefly discussed in \cite{wen2021attacks}. 
\end{itemize}

\section{Conclusion}
\label{Conclusion}
In recent years, blockchain has seen a dramatic increase in popularity, driven by the rise of decentralized systems and its widespread applications across various fields like cryptocurrencies, banking sectors, crypto-assets, IOT and health services. Popularity comes with a considerable cost of security threat \& attacks on established systems using blockchain. Over the years, the losses have been mounting \cite{Security_2024}. These exposed vulnerabilities weaken the legitimacy of the blockchain as a decentralized system and demands to be addressed \& studied thoroughly. In our study, to evaluate the scenarios of security threats \& impacts,  we used the four layers: network, consensus, application, and meta-application. For every studied attack, we discussed: i) attacker's motivation \& vulnerabilities that might lead to an attack, ii) the attack strategy for each attack is represented visually, iii) the conditions for the attack and the possible attack outcome, iv) the possibility of leading to another attack and v) the practicality of the attack \& possible countermeasures. Through detailed analysis, it is evident that attacks can transcend layers, with vulnerabilities in one layer potentially leading to exploits in another. Furthermore, our findings are presented in tabular formats and demonstrate that a single attack can span multiple layers, depending on the targeted attributes of each layer, thereby amplifying the complexity of securing blockchain systems.

In conclusion, we believe that no system is completely secure. Therefore, additional research is required to develop a more secure and functional architecture for blockchain-based systems. This survey aims to provide valuable insight into different aspects of the examined attacks, as well as to highlight the intricate connectivity across layers under various attack scenarios. This survey can serve as an effective guide to limit adversarial activity across blockchain layers, leading to intriguing outcomes based on future research in this field.
\newcolumntype{C}[1]{>{\centering\arraybackslash}p{#1}}

{ \small 
\begin{landscape}
\captionsetup{type=table,justification = centering}
\caption{Comparison with previous works}
\label{tab:comparison_of_works}

\begin{longtable}{@{}|C{2.5cm}|C{1.2cm}|C{1.2cm}|C{1.7cm}|C{1.4cm}|C{1.3cm}|C{1.2cm}|p{6.3cm}|@{}}
\toprule
\rowcolor[HTML]{F0EBEB} 
{\color[HTML]{000000} \textbf{Reference}} &
  {\color[HTML]{000000} \textbf{Number of Attacks Covered}} &
  {\color[HTML]{000000} \textbf{Number of Layers}} &
  {\color[HTML]{000000} \textbf{Inter-relation of Attacks Between layers}} &
  {\color[HTML]{000000} \textbf{Mitigation Techniques}} &
  {\color[HTML]{000000} \textbf{Attacker's Perspective Analysis}} &
  {\color[HTML]{000000} \textbf{Layer Vulnerability Analysis}} &
  {\color[HTML]{000000} \textbf{Remarks}} \endhead  \midrule
  \multirow{3}{*}{Homoliak \textit{et al.}\cite{homoliak2020security}} &
  \multirow{3}{*}{$\approx$ 21} &
  \multirow{3}{*}{4} &
  \multirow{3}{*}{\Circle} &
  \multirow{3}{*}{\CIRCLE} &
  \multirow{3}{*}{\Circle} &
  \multirow{3}{*}{\CIRCLE} &
   \textbf{1.} Focused on SRA of blockchain systems. \newline
   \textbf{2.} Used 4 layers to categorize security threats and vulnerabilities.
  \\ \midrule
\rowcolor[HTML]{EEEEEE} 
\multirow{4}{*}{Saad \textit{et al.}\cite{saad2020exploring}} &
   \multirow{4}{*}{22}&
   \multirow{4}{*}{0} &
  \multirow{4}{*}{\Circle} &
  \multirow{4}{*}{\CIRCLE} &
  \multirow{4}{*}{\Circle} &
  \multirow{4}{*}{\Circle} &
  \textbf{1.} Emphasize on blockchain attack surface.\newline
   \textbf{2.} Outlined effective defense measures.\newline
   \textbf{3.} Explored cryptographic constructions, distributed system
    architecture, and application of blockchain.
   \\ \midrule
Guggenberger \textit{et al.}\cite{guggenberger2021structured} &
  \multirow{2}{*}{87} &
   \multirow{2}{*}{0} &
  \multirow{2}{*}{\Circle} &
  \multirow{2}{*}{\Circle} &
  \multirow{2}{*}{\Circle} &
  \multirow{2}{*}{\Circle} &
  \multirow{2}{*}{\textbf{1.} Structured attacks using AT notation.}
  \\ \midrule
  \rowcolor[HTML]{EEEEEE} 
\multirow{4}{*}{Li \textit{et al.}\cite{li2020survey}} &
 \multirow{4}{*}{ 6} &
  \multirow{4}{*}{0} &
  \multirow{4}{*}{\Circle} &
  \multirow{4}{*}{\RIGHTcircle} &
  \multirow{4}{*}{\Circle }&
  \multirow{4}{*}{\Circle }&
  \textbf{1.} Discussed security vulnerabilities related to blockchain.\newline
   \textbf{2.} Reviewed security enhancement solutions for blockchain.\\ \midrule
\multirow{3}{*}{Chen \textit{et al.}\cite{chen2020survey}} &
  \multirow{3}{*}{26} &
   \multirow{3}{*}{4} &
  \multirow{3}{*}{\Circle} &
  \multirow{3}{*}{\RIGHTcircle} &
  \multirow{3}{*}{\Circle} &
  \multirow{3}{*}{\CIRCLE} &
  \textbf{1.} Focused on Ethereum platform security.\newline
   \textbf{2.} Considered three perspectives- vulnerabilities, attacks and defense.\\ \midrule
  \rowcolor[HTML]{EEEEEE} 
\multirow{2}{*}{Wen \textit{et al.}\cite{wen2021attacks}} &
  \multirow{2}{*}{17} &
  \multirow{2}{*}{ 6} &
  \multirow{2}{*}{\Circle} &
  \multirow{2}{*}{\CIRCLE }&
  \multirow{2}{*}{\RIGHTcircle} &
  \multirow{2}{*}{\Circle} &
  \textbf{1.} Focused on attacks on blockchain system and their countermeasures.\\ \midrule
   \multirow{2}{*}{Moubarak \textit{et al.} \cite{moubarak2018blockchain}} &
  \multirow{2}{*}{7} &
  \multirow{2}{*}{0} &
  \multirow{2}{*}{\Circle} &
  \multirow{2}{*}{\RIGHTcircle} &
  \multirow{2}{*}{\Circle} &
  \multirow{2}{*}{\Circle} &
  \textbf{1.} Evaluated blockchain security.\newline
   \textbf{2.} Identified flaws that may lead to attacks.
   \\ \midrule 
   \multirow{2}{*}{Anita \textit{et al.} \cite{ anita2019blockchain}} &
  \multirow{2}{*}{17} &
  \multirow{2}{*}{0} &
  \multirow{2}{*}{\Circle} &
  \multirow{2}{*}{\CIRCLE} &
  \multirow{2}{*}{\Circle} &
  \multirow{2}{*}{\Circle} &
  \textbf{1.} 1. Presented taxonomy of security threats related to blockchain systems.
   \\ \midrule 
    \multirow{3}{*}{Chen \textit{et al.} \cite{chen2022survey}} &
  \multirow{3}{*}{11} &
  \multirow{3}{*}{0} &
  \multirow{3}{*}{\Circle} &
  \multirow{3}{*}{\CIRCLE} &
  \multirow{3}{*}{\Circle} &
  \multirow{3}{*}{\Circle} &
  \textbf{1.} Reviewed the attack and defense methods of the blockchain.\newline
   \textbf{2.} Categozied attacks into 3 different groups.
   \\ \midrule 
   \multirow{3}{*}{Zhang \textit{et al.} \cite{zhang2019security}} &
  \multirow{3}{*}{4} &
  \multirow{3}{*}{6} &
  \multirow{3}{*}{\Circle} &
  \multirow{3}{*}{\Circle} &
  \multirow{3}{*}{\Circle} &
  \multirow{3}{*}{\Circle} &
  \textbf{1.} Concentrated on the security and privacy aspects and mechanisms of blockchain.\newline
   \textbf{2.} Compared different types of consensus mechanism.
   \\ \midrule 
    \multirow{4}{*}{This work} &
  \multirow{4}{*}{23} &
  \multirow{4}{*}{4} &
  \multirow{4}{*}{\CIRCLE} &
  \multirow{4}{*}{\CIRCLE} &
  \multirow{4}{*}{\CIRCLE} &
  \multirow{4}{*}{\CIRCLE} &
  \textbf{1.} Focused on Layer-based attack analysis.\newline
   \textbf{2.} Covered attacker's perspective analysis in detail.\newline
   \textbf{3.} Covered layer vulnerability analysis.\newline
   \textbf{4.} Covered mitigation techniques.\\ \midrule
\end{longtable}
\end{landscape}
}
\bibliographystyle{ACM-Reference-Format}
\bibliography{main}

\end{document}